# First Detection and Modeling of Spatially Resolved Lyα in TW Hya

Seok-Jun Chang,[1]⋆ Nicole Arulanantham,[2] Max Gronke,[1] Gregory J. Herczeg,[3] Edwin A. Bergin,[4]

[1] *Max-Planck-Institut für Astrophysik, Karl-Schwarzschild-Straße 1, 85748 Garching b. München, Germany*
[2] *Space Telescope Science Institute, Baltimore, USA*
[3] *Kavli Institute for Astronomy and Astrophysics, Peking University, Peking, China*
[4] *Department of Astronomy, University of Michigan, Michigan, USA*

## ABSTRACT

Lyman-α (Lyα) is the strongest emission line in the accretion-generated UV spectra from T-Tauri stars and, as such, plays a critical role in regulating chemistry within the surrounding protoplanetary disks. Due to its resonant nature, the scattering of Lyα photons along the line-of-sight encodes information about the physical properties of the intervening H I medium. In this work, we present the first spatially resolved spectral images of Lyα emission across a protoplanetary disk in the iconic face-on T-Tauri star TW Hya, observed with HST-STIS at spatial offsets 0″, ±0.2″, and ±0.4″. To comprehensively interpret these Lyα spectra, we utilize a 3D Monte-Carlo Lyα radiative transfer simulation considering the H I wind and protoplanetary disk. From the simulation, we constrain the wind's properties: the H I column density $\sim 10^{20}$ cm$^{-2}$ and the outflow velocity $\sim 200$ km s$^{-1}$. Our findings indicate that successfully interpreting the observed spectra necessitates scattering contributions in the H I layer within the disk. Furthermore, to explore the effect of Lyα radiative transfer on protoplanetary disk chemistry, we compute the radiation field within the scattering medium and reveal that the wind reflection causes more Lyα photons to penetrate the disk. Our results show the necessity of spatially resolved Lyα observations of a broad range of targets, which will decode the complex interactions between the winds, protoplanetary disks, and surrounding environments.

## 1 INTRODUCTION

The accretion disks around young stellar objects (YSOs) are the birthplaces of planetary systems, providing reservoirs of icy planetesimals and warm gas and dust from which cores and atmospheres are assembled. Observations indicate that the thermal structure and chemical composition of molecular gas in surface layers of these protoplanetary disks are influenced in part by UV radiation fields, generated both internally by the central YSOs (see, e.g., Pascucci et al. 2009; Miotello et al. 2019; Bergner et al. 2019, 2021; Banzatti et al. 2020; Teague & Loomis 2020) and externally by neighboring massive stars (see e.g., van Terwisga & Hacar 2023; Maucó et al. 2023). The UV emission from many YSOs is enhanced above the standard stellar blackbody emission by continuum and line emission from the hot accretion shocks (Johns-Krull et al. 2000). Generally, the emission of hydrogen Lyα transition dominates the UV fluxes associated with T Tauri stars (e.g. Schindhelm et al. 2012; France et al. 2014), the subset of YSOs less massive than $\sim 2\,M_\odot$.

Lyα emission lines observed from T Tauri stars show broad line wings consistent with resonant scattering through a surrounding H I medium (Arulanantham et al. 2023). This includes the surface layers of protoplanetary disks ($n < 10^6$ cm$^{-3}$; Bethell & Bergin 2011), infalling H I associated with accretion flows, and magnetically or photoevaporatively driven winds (Kurosawa et al. 2006; Arulanantham et al. 2021). Models predict that Lyα scattering through the disk will enhance the UV flux in deeper, radially extended surface layers that are optically thick to photons from the FUV continuum (Bergin et al. 2003; Bethell & Bergin 2011). However, the fraction of Lyα that escapes the accretion columns and outflowing material to reach the outer disk has yet to be measured from observations. While the Lyα illuminating hot $H_2$ in the inner disk is positively correlated with observed accretion tracers (France et al. 2023), variance in disk flaring, outflow properties, and resonant scattering at large radii may complicate the relationship between emission line fluxes, spatial distributions of molecular gas, and accretion luminosities (see, e.g., Bergner et al. 2021).

Protoplanetary disks that are viewed face-on offer an ideal environment in which to explore the impact of Lyα scattering. In this work, we examine *Hubble Space Telescope (HST)* spectral imaging of the face-on disk around the roughly solar mass T Tauri star TW Hya, at 60 pc (Gaia Collaboration et al. 2023) one of the closest accreting YSOs, to search for Lyα emission extending beyond the central accretion shocks. In Section 2, we explain the target and observational setup details. In Section 3, we show the observational data of spatially resolved Lyα in TW Hya via the HST 2D spectra with various spatial offsets. In Section 4, we analyze the observed spectra through modeling adopting 3D Monte-Carlo Lyα radiative transfer, including scattering through an inner disk wind that was observed in [O I] emission (Fang et al. 2023). The models constrain the physical properties of the wind and the disk required to reproduce the Lyα spectrum on the star and the spatial distribution of Lyα. In Section 5, we discuss the effect of Lyα radiative transfer in T-Tauri stars through the internal Ly$\alpha$ radiation field within the wind and disk. In addition, we explore the dependence of Ly$\alpha$ spectra on the observing direction. Section 6 summarizes our main conclusions.

⋆ E-mail: sjchang@mpa-garching.mpg.de

## 2 TARGET & OBSERVATIONS

We acquired spectral imaging of TW Hya, using the Space Telescope Imaging Spectrograph onboard the *Hubble Space Telescope* (*HST*-STIS), to explore the spatial distribution of resonantly scattered Ly$\alpha$ in a protoplanetary disk environment for the first time (Program ID: 11607; PI: T. Bethell). As one of the closest young accretion disk systems ($d = 60.0^{+0.4}_{-0.1}$ pc; Gaia Collaboration 2016, 2023), TW Hya has been the subject of numerous observational studies from X-ray through sub-mm wavelengths. Its dust disk is highly structured, with gaps at $r \sim 1, 25.62, 31.5, 41.64$, and 48 au (Huang et al. 2018a). Moving shadows cast across its surface indicate that two misaligned inner disks are present at $r \sim 5$ and $r \sim 6$ au (Teague et al. 2022; Debes et al. 2023).

In general, the winds from classical T Tauri stars have several different components: a fast wind or jet from near the star-disk interaction region (e.g. Edwards et al. 2006; Banzatti et al. 2019; Xu et al. 2021), a slower MHD wind (e.g. Wang et al. 2019; Fang et al. 2023), and a photoevaporative wind (e.g. Gorti & Hollenbach 2008; Ercolano & Owen 2010). These structures are inferred from P Cygni absorption profiles and forbidden emission lines, though often the geometry limits the components seen for any individual star (see review by Pascucci et al. 2023). TW Hya itself exhibits strong wind absorption at velocities from 100–200 km s$^{-1}$ (Dupree et al. 2005; Johns-Krull & Herczeg 2007) and narrow emission in [O I] dominated by either an MHD or photoevaporative wind within $r < 1$ au (Fang et al. 2023; Rab et al. 2023). The star has maintained a stable accretion rate of $2.5 \times 10^{-9} M_\odot$ yr$^{-1}$ for at least the past 25 years (Herczeg et al. 2023), with sporadic X-ray and FUV flares (e.g. Brickhouse et al. 2012; Hinton et al. 2022). Searches for accreting protoplanets within the disk have not yet yielded any candidate companions (Huélamo et al. 2022; Sicilia-Aguilar et al. 2023).

The central Ly$\alpha$ spectrum from TW Hya was first presented in Herczeg et al. (2004), who found that this single emission line comprised $\sim$85% of the system's total FUV flux. Models of the H I absorption feature superimposed on the emission line are consistent with very little interstellar reddening ($E(B − V) < 0.01$; Herczeg et al. 2004; McJunkin et al. 2014). The broad emission line width encompasses photodissociation cross sections of HCN and H$_2$O (Mordaunt et al. 1994; Bergin et al. 2003; van Harrevelt & van Hemert 2008; Heays et al. 2017), neither of which have been detected in the warm inner disk (Najita et al. 2010), despite a clear detection of HCN in the outer disk (Kastner et al. 1997).

To map the spatial distribution of Ly$\alpha$ across the disk surface, additional *HST*-STIS observations were acquired with the $52'' \times 0.2''$ long slit, using the FUV-MAMA and the G140M grating ($\lambda_0 = 1222$ Å; $R \sim 10,000$). While a larger slit width in the dispersion direction would have allowed for higher throughput in the Ly$\alpha$ emission line wings than the 0.2$''$ slit width selected for these observations, the narrower entrance aperture minimizes the contamination from scattering within the detector and G140M grating. The slit was stepped across the disk in increments of 0.2$''$ to image the Ly$\alpha$ flux between $\pm 0.5''$ of the star (see Figure 9). At the source distance of 60 pc, the images span a radial extent of $\pm 24$ au with a plate scale of $\sim$1.7 au pixel$^{-1}$ (0.029$''$ pixel$^{-1}$).

PSF subtraction is required to separate the spatially extended Ly$\alpha$ emission from the stellar/accretion shock contribution in the off-star spectral images since fringe patterns can be produced by Airy rings entering the slit when adjacent point sources are not centered in the cross-dispersion (spatial) direction of the slit. In order to characterize this instrument response to the stellar Ly$\alpha$ flux at offset positions of $\pm 0.2''$ and $\pm 0.4''$, a nearby white dwarf, WD-1056-384, was observed with identical configurations and slit positions as those used to observe TW Hya. The white dwarf produces strong continuum emission across the entire UV bandpass, providing an estimate of the fringe pattern from a bright, stellar point source with no extended circumstellar material that is deliberately not centered in the slit. Spectral images from the white dwarf will, therefore, demonstrate the extent to which the PSF spreads the stellar flux to larger angular separations, where it may mask any Ly$\alpha$ emission originating from the circumstellar disk surrounding TW Hya.

The spectral image centered on WD-1056-384 shows a bright continuum in the dispersion direction, concentrated between $\pm 0.1''$ in the cross-dispersion direction. The spectrum diminishes near Ly$\alpha$ line center (1215-1216 Å), where the white dwarf has a strong absorption feature. Geocoronal emission fills the slit at the line center in all five images, with a surface brightness of $\sim 10^{-12}$ erg s$^{-1}$ cm$^{-2}$ arcsec$^{-2}$. Significant extended emission is seen in the images taken at $\pm 0.2''$, at a surface brightness approximately two orders of magnitude lower than observed in the image centered on the white dwarf. This diffuse flux spans $\sim \pm 0.4''$ along the slit but does not show a prominent fringe pattern. We scale the spectral images from WD-1056-384 to the peak surface brightness of TW Hya at each on-source and offset position and subtract the resulting PSF "model", leaving behind the Ly$\alpha$ emission generated at the source. Additional details are discussed in Appendix A.

From each PSF-subtracted spectral image of TW Hya, we extracted 1-D spectra at different positions along the slit to investigate the shape of the Ly$\alpha$ emission line profiles as a function of distance from the central star. This was performed by summing 0.25$''$ (15 au) segments of the flux-calibrated spectral image arrays along the length of the slit. The integrated surface brightnesses were then multiplied by the slit width (0.2$''$) and the plate scale in the spatial direction (0.029$''$ pixel$^{-1}$). Since the distribution of Ly$\alpha$ flux along the slit is not spatially uniform, we also account for slit losses and scale by the point source aperture throughput and slit height [1].

## 3 OBSERVATIONAL RESULTS

Figure 1 shows the Ly$\alpha$ spectral images of TW Hya, acquired with the slit centered on the star and at offsets of $\pm 0.2''$ and $\pm 0.4''$. In each image, distinct red and blue peaks are detected with a dark, outflow-absorbed central region at the line center. The image centered on the star shows a spatially concentrated distribution of flux, with a peak surface brightness of $\sim 6 \times 10^{-10}$ erg s$^{-1}$ cm$^{-2}$ arcsec$^{-2}$ at $\sim$250 km s$^{-1}$, which corresponds to the red vertical dashed line in Figure 1. The fainter peak on the blue side of the emission line is separated from the red peak by 3.96 Å (or $\sim$980 km s$^{-1}$). The spatial distribution of the Ly$\alpha$ flux is broader in the slit positions further from the star, showing a maximum surface brightness of $\sim 0.4 - 0.7 \times 10^{-10}$ erg s$^{-1}$ cm$^{-2}$ arcsec$^{-2}$ in the spectra acquired at offsets of $\pm 0.2$". At $\pm 0.4$", the maximum surface brightness decreases to $\sim 0.1 \times 10^{-10}$ erg s$^{-1}$ cm$^{-2}$ arcsec$^{-2}$. The separation between blue and red peaks

---

[1] See Section 5.4 of the STIS Data Handbook: https://hst-docs.stsci.edu/stisdhb/chapter-5-stis-data-analysis/5-4-working-with-spectral-images

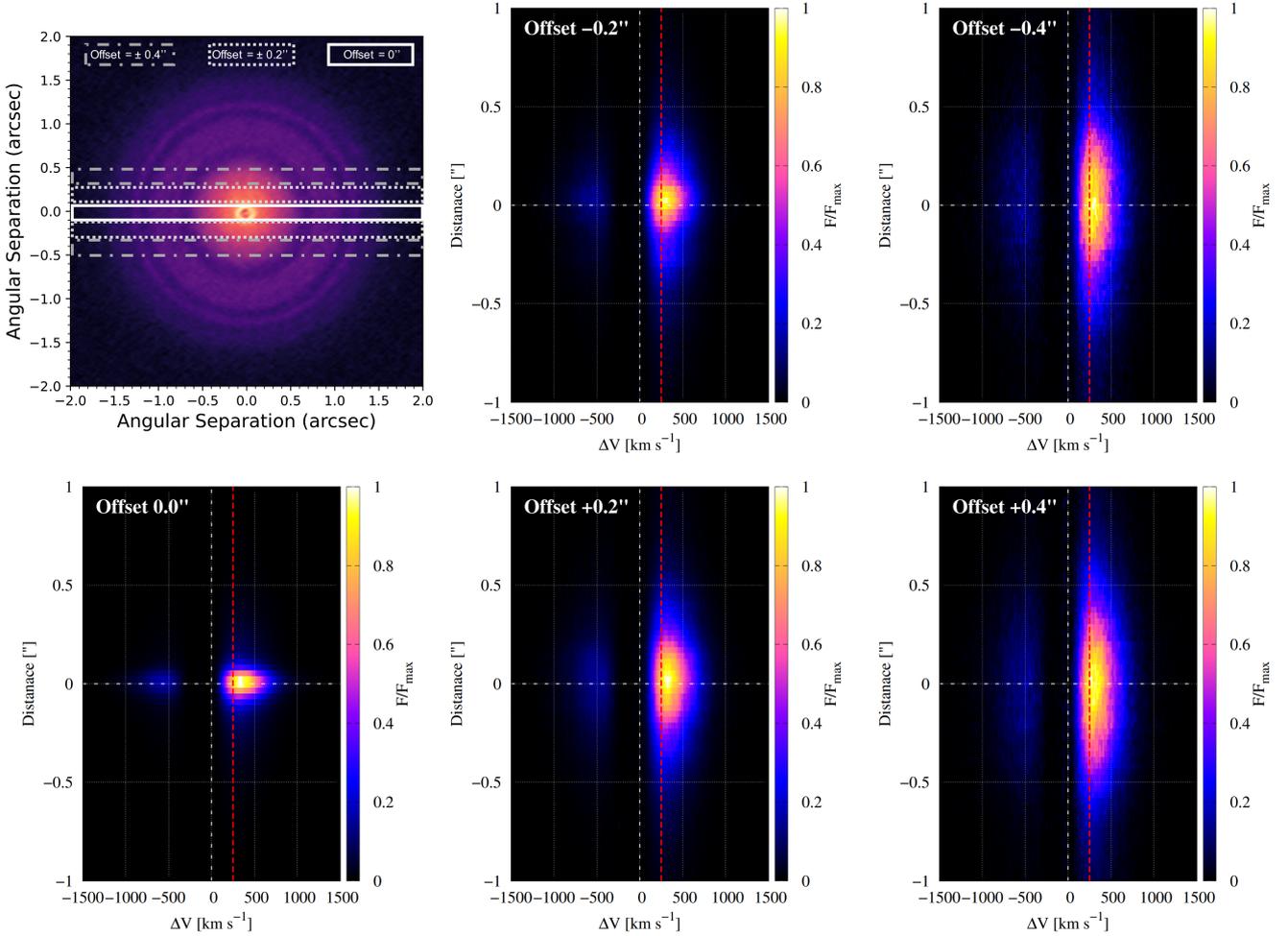

**Figure 1.** Observed 2D Lyα spectra of Lyα obtained by HST for various offsets 0″, ±0.2″, and ±0.4″. The left top panel shows the 290 GHz continuum image from [Huang et al. (2018a)](#), where the STIS slit positions are sketched. The 2D spectra are normalized by dividing by the maximum value. All spectral profiles are composed of weak-blue and strong-red peaks in the Doppler space. When the 2D spectrum is far from the center (±0.4″), it is spatially extended.

also decreases to 3.71 Å (∼ 915 km s⁻¹) at ±0.2″ and 3.63 Å (∼ 896 km s⁻¹), at ±0.4″.

Figure 2 shows the 1D spectra of Lyα extracted from the 2D spectral images in Figure 1 and the integrated Lyα spectra of each 2D spectrum. In the top panel, we compare the 1-D spectrum acquired on the source to spectra acquired at ±0.875″ = ±50 au. The emission line profile centered on the star has broad emission line wings extending to ±2000 km s⁻¹ and an absorption depth of roughly two orders of magnitude. At ±50 au, the absorption depth shrinks to less than an order of magnitude. The bottom panel displays the integrated Lyα spectra for various spatial offsets. The blue and red wings in the offset integrated spectra are slightly weaker than those of the spectrum on the star, as expected for line profiles that are produced by Lyα scattering in a spatially extended H I medium.

The N V resonance doublet at 1238.82, 1242.80 Å is also detected in the spectra of both TW Hya and WD-1056-384. Unlike the Lyα emission, the N V features are not spatially extended and do not show significant surface brightness outside of ±0.4″ (see Figure 3). N V is scattered by dust or by atomic hydrogen when the H I column density is higher than ∼ 10²² cm⁻²; the cross section of Rayleigh scattering with atomic hydrogen at 1240 Å is ∼ 7.5 × 10⁻²³cm² ([Lee 2013](#); [Chang et al. 2017](#)). Our observations are then consistent with a profile that originates at the accretion shocks ([Ardila et al. 2013](#)), with little scattering by dust grains across the disk. If the layer of H I is depleted of small grains, Lyα photons can more easily penetrate the molecular gas surface ([Bethell & Bergin 2011](#)). In the following sections, we use the *HST*-STIS observations to constrain simulations of Lyα resonant scattering through a protoplanetary disk and wind and discuss the implications for chemistry in the era of planet formation.

## 4 SIMULATION & ANALYSIS

Previous analyses of Lyα emission from T Tauri stars have exclusively focused on the 1-D spectral profiles. For instance, [Herczeg et al. (2004)](#) modeled the spectrum from TW Hya by superimposing a broad Gaussian emission component and a Voigt profile to represent absorption by the H I medium ($N_{HI} ∼ 10^{19}$ cm⁻²), following methods developed by [Wood et al. (2002)](#). This technique was later applied to a larger sample of *HST*-COS spectra from YSOs to measure the total column densities of intervening H I ([McJunkin et al. 2014](#)). The Lyα spectrum reaching the molecular gas disk inside $r < 10$ au was also reconstructed using the fluxes from UV-fluorescent H₂ emission lines (Herczeg et al. 2004; Schindhelm et al. 2012). Arulanantham et al. (2023) included resonant scattering to reproduce the line profiles by utilizing Lyα radiative transfer in a simple shell model (Ahn et al. 2001; Gronke et al. 2015). The shell model comprises a central source surrounded by a thin H I layer with constant radial velocity, which was able to reproduce the blue- or red-enhanced Lyα peaks corresponding to intervening accretion flows or outflowing winds, respectively.

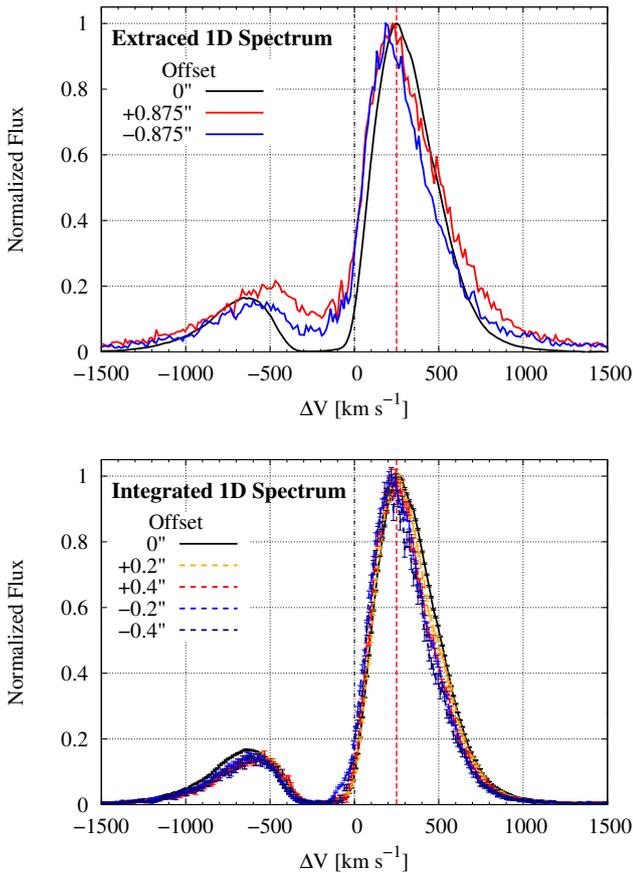

**Figure 2.** 1D spectra of Lyα from TW Hya, as observed with *HST*-STIS. *Top:* The Lyα spectra extracted from the 2D spectral image centered on the star, at separations of 0″ (black line) and +/-0.875″ (red and blue lines) along the slit and within widths of 0.25″. While the line profiles from +/-0.875″ have similar shapes in the redder peaks, they have shallower absorption features near the systemic velocity blueward of the line center. *Bottom:* The 1D spectra were measured by integrating over the 2D spectral images centered on the star (black, solid line) and at offsets of ±0.2″ and ±0.4″ (red and blue dashed lines). The spectra from the offset images show slightly weaker red and blue wings than the on-source spectrum.

However, these prior works did not consider the spatial distribution of Lyα, which we have detected for the first time in a protoplanetary disk with *HST*-STIS. Significant Lyα emission is observed at separations up to ±2″ (see Figure 1), well outside the accretion shocks where the photons originate. The diffuse UV flux may influence chemistry at large radial distances and vertical depths within the disk where FUV continuum photons are not able to penetrate (Bergin et al. 2003; Bethell & Bergin 2011); however, the computational challenge of integrating models of Lyα radiative transfer, protoplanetary disk chemistry, and magnetothermal winds has made this difficult to explore without additional observational constraints. Our study investigates the Lyα spectrum and surface brightness distribution from TW Hya through 3D radiative transfer simulation that describes the propagation of photons through the protoplanetary disk and wind. The following sections will describe the geometries of the models, along with our simplifying assumptions.

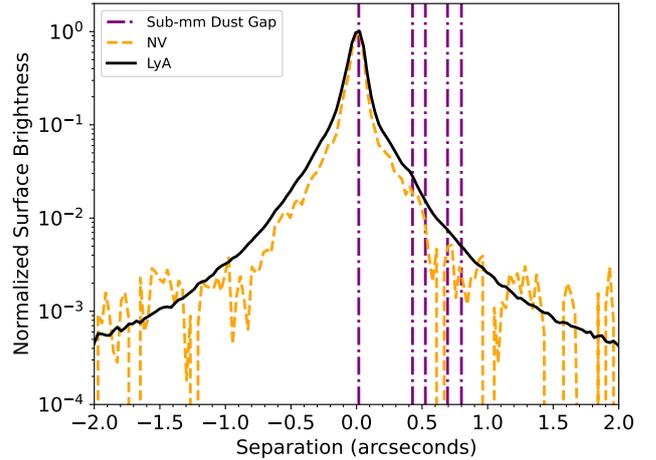

**Figure 3.** Radial distribution of Lyα flux from TW Hya, measured from *HST*-STIS long-slit spectroscopy, with vertical lines marking the sub-mm dust gaps identified by Huang et al. 2018a. The distribution is brightest in the spectrum acquired on the source (black, solid line), and the profile is narrow between ±0.5″. The peak surface brightness decreases by order of magnitude at an offset of 0.2″ and two orders of magnitude at 0.4″, and the profiles become wider. No structure is observed in the Lyα profile, indicating that the extended emission is produced by resonant scattering in a smooth H I layer. The orange dashed line represents the radial distribution of N v flux less extended than that of Lyα.

### 4.1 Monte Carlo simulation for Lyα radiative transfer

We use a Lyα radiative transfer (RT) simulation developed by Chang et al. (2023), based on *LaRT* (Seon & Kim 2020; Seon et al. 2022). This RT simulation is capable of considering arbitrary 3D scattering geometries with kinematics in Cartesian coordinates, making it an ideal first approach for examining the propagation of Lyα photons through the multi-phase H I medium surrounding T Tauri stars (see, e.g., McJunkin et al. 2014). The 3D scattering geometry in the simulation is composed of $300^3$ cells.

Near-infrared scattered light observations of TW Hya identify a population of sub-$\mu m$ grains that are coupled to the gas layer (van Boekel et al. 2017), with a grain size distribution consistent with the ISM dust models of Mathis et al. 1977. Within the H I layer, we, therefore, adopt the dust model from Draine (2003a,b) to describe dust absorption and scattering properties of ISM-like grains. The photons in the simulation are characterized by their position **r**, direction $\hat{\mathbf{k}}$, wavelength $\lambda$, and polarization vector (Stokes vector) $\mathbf{S} = (I, Q, U, V)$. The Monte Carlo simulation proceeds through the following steps: (1) generating intrinsic Lyα photons from the emission region, (2) determining the next scattering location, (3) calculating the information of scattered photons, and (4) repeating steps (2)-(3) until the photon escapes the entire grid (see Chang et al. (2023) for additional details). The simulation generates $10^6$ photons for each parameter set. Additionally, we adopt the peeling-off tech-nique [2] (Yusef-Zadeh et al. 1984; Laursen & Sommer-Larsen 2007). It allows us to obtain the Lyα spectrum and surface brightness along specific lines of sight, which we use to compare to the observations of TW Hya and to estimate the spatial distribution of Lyα in more inclined disks.

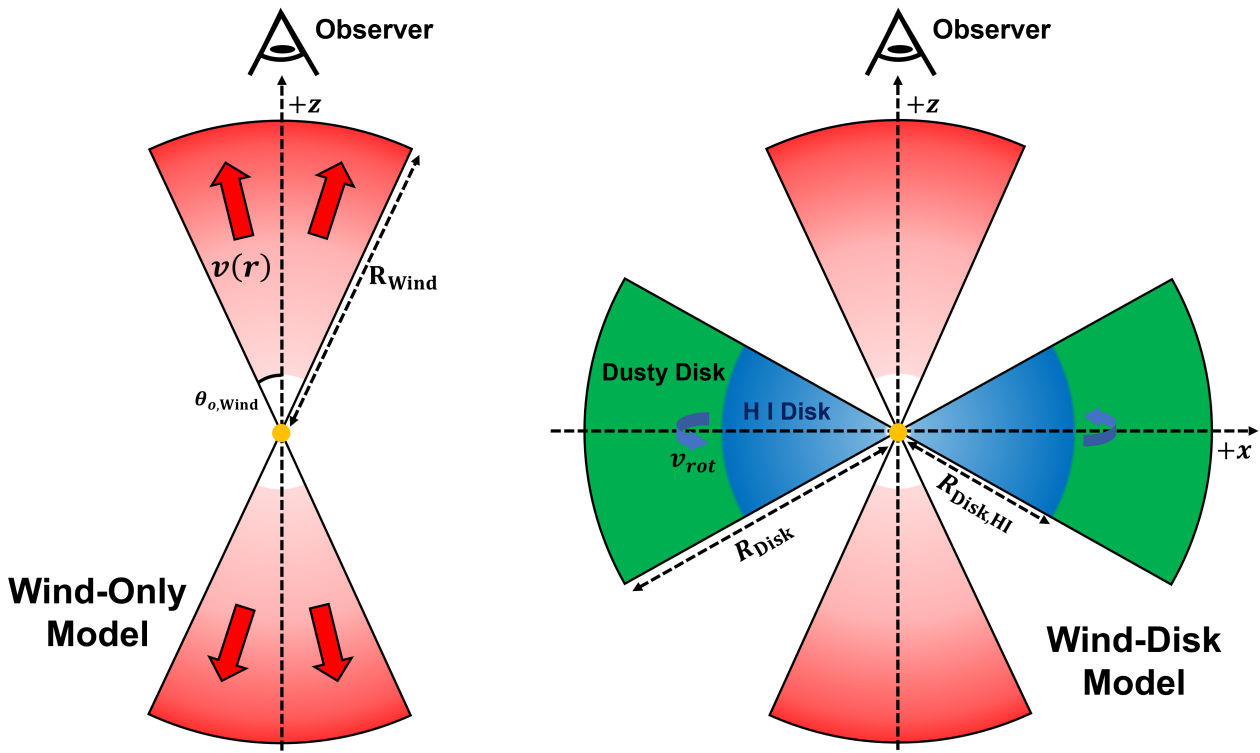

**Figure 4.** Schematic illustrations of the two geometries: the 'Wind-Only Model' (left) and the 'Wind-Disk Model' (right). *Left:* The Wind-Only Model comprises the central Lyα source (orange) and the bipolar outflowing wind (red). The bipolar wind is characterized by a radius $R_{\text{Wind}}$, an opening angle $\theta_{o,\text{Wind}}$, and a radial outflow velocity $v(r)$. *Right:* The Wind-Disk Model incorporates an additional equatorial disk, consisting of an inner H I disk (blue) and an outer dusty disk (green). The disk is characterized by a disk radius $R_{\text{Disk}}$ and the inner H I disk radius $R_{\text{Disk,HI}}$. The dusty disk has a dust optical depth $\gg 1$; Lyα is impossible to pass through it. The line of sight is fixed in the +$z$ direction, as TW Hya is observed as a face-on T Tauri star.

It is important to note that the simplified geometry presented here implies that the parameters should not be interpreted literally. Instead, this first exploration shows how the physical properties of a H I medium are imprinted on the observable Lyα emission features. In the following section, we introduce two types of our simplified geometries: the 'wind-only model', which explains how the Lyα profile is formed, and the 'wind-disk model', which explains how the Lyα emission spatially extended from the central star to the protoplanetary disk.

### 4.2 Geometry

Our model consists of three main components: a central source emitting Lyα photons, a bipolar wind, and a dusty rotating disk. Figure 4

provides schematic illustrations of the two models: the wind-only and the wind-disk models. Since the protoplanetary disk surrounding TW Hya is face-on, the line of sight is set in the +$z$ direction. Initially, we analyze the Lyα spectrum through the wind-only model to derive the physical properties of the wind. Subsequently, we incorporate a simplified protoplanetary disk model, which includes an H I layer extending to a H I disk radius $R_{\text{Disk,HI}}$ and a dusty H I-free region at larger radii. We note that the volume of gas contained within $R_{\text{Disk,HI}}$ is more likely distributed in a thin surface layer that spans the full radial extent of the disk (see, e.g., Huang et al. 2018b). However, this approximation allows us to determine to first order whether scattering through the outflowing wind or across the disk contributes more to the spatially extended Lyα emission observed with *HST*-STIS. In the following sections, we describe the details and physical motivations of the two models.

#### 4.2.1 Wind-Only Model

The presence of a stellar wind in TW Hya has long been inferred from He I 10830 Å observations (Dupree et al. 2014), along with a photoevaporative wind detected in [Ne II] 12.81 μm (Pascucci et al. 2011) and an MHD wind that was spatially resolved in [O I] 6300 Å emission (Fang et al. 2023). The wind adopted in our models is most similar to the stellar and MHD winds, which the Lyα photons would encounter first upon emerging from the central emitting region. The Lyα 1D spectral profile from TW Hya is dominated by scattering

---

through these two components and is also most consistent with the T Tauri profiles (Arulanantham et al. 2023).

The wind-only model, depicted in the left panel of Figure 4, comprises a central Lyα source (accretion shock) emitting photons with an intrinsic line width $\sigma_{src}$, and a bipolar wind characterized by an opening angle $\theta_{o,Wind}$, H I column density $N_{HI,Wind}$, wind radius $R_{Wind}$, and temperature $T_{Wind}$. However, we note that these wind properties are not yet well constrained by observations. For Class II sources, Pascucci et al. (2023) reported that the half-opening angles of high-velocity jet and low-velocity wind emissions range from $1 - 5°$ and $25 - 35°$, respectively. To capture the full extent of H I across the disk, we assume an opening angle of the wind $\theta_{o,Wind} = 30°$. We also assume $T_{Wind} \leq 10^4$ K, representing thermal motion $v_{th} = \sqrt{2kT_{Wind}/m_H}$, where $m_H$ is a hydrogen mass.

The velocity structure of winds has been measured across the surfaces of protoplanetary disks, with multi-wavelength gas emission lines tracing a shell-like structure (see, e.g., Pascucci et al. 2023). However, few constraints are available for the velocity gradient in the vertical direction ($+z$ in our models). The He I 10830 Å spectrum from TW Hya shows a broad absorption feature in the blueward with velocities ranging from $-400$ to $-100$ km s$^{-1}$, indicating that multiple velocity components contribute to the profile of even a single wind tracer (Dupree et al. 2014). To accommodate this gradient, along the radial direction from the central source, we assume a Hubble-like outflow velocity $v(r)$ given by

$$v(r) = v_{exp}\left(\frac{r}{R_{Wind}}\right), \qquad (1)$$

where $v_{exp}$ is the maximum expansion velocity. Since the density profile of the winds is also observationally unconstrained, we assume a uniform H I number density such that

$$n_{HI} = \frac{N_{HI,Wind}}{R_{Wind} - R_{i,Wind}}, \qquad (2)$$

where $R_{i,Wind}$ represents the inner radius of the wind. The H I column density of the wind $N_{HI,Wind}$ is estimated to be around $\sim 10^{19-20}$ cm$^{-2}$, based on the simple absorption model fit to the 1D Lyα spectral profile (Herczeg et al. 2004). We account for the ionization of the inner wind by UV radiation from the central star by setting $R_{i,Wind} = 0.2R_{Wind}$. We will discuss the best-fit model and compare the *HST*-STIS spectra to the wind-only model in Section 4.3.

### 4.2.2  Wind-Disk Model

The wind-disk model, shown in the right panel of Figure 4, extends the wind-only model by including an equatorial protoplanetary disk component. The H I region at the surface of the protoplanetary disk (Bergin et al. 2003) induces the spatial diffusion of Lyα via scattering (Chang et al. 2023). As our HST data demonstrates in Figure 1, this H I component is crucial to accurately reproduce the Lyα radiation seen by the molecular gas disk. Therefore, the equatorial disk in the right panel of Figure 4 consists of two types of regions: an inner H I region (blue) with a radius of $r < R_{Disk,HI}$ and an outer dusty disk (green) beyond $R_{Disk,HI}$ with a dust optical depth of $\tau_d = 10$ per 1 au. In the inner H I disk, a H I number density $n_{HI,Disk}$ is fixed at $10^7$ cm$^{-3}$, given that H I number densities in protoplanetary disks are expected to be $> 10^6$ cm$^{-3}$ (Bethell & Bergin 2011). We fix the outer radius of the disk $R_{Disk}$ at 100 au, based on estimates from SPHERE and observations of the sub-$\mu$m dust distribution in TW Hya (van Boekel et al. 2017). This limit also coincides with the steep decline in the $^{12}$CO surface brightness profile observed at $r \sim 90$ au with ALMA (Huang et al. 2018b). The inner radius of the disk is set at $R_{Disk}/100$ due to the resolution of the grid in the RT simulation.

To represent the vertical disk structure $H_{Disk}$ as a function of radius, we assume

$$H_{Disk}(r) = R_{Disk} \tan 30° \left(\frac{r}{R_{Disk}}\right)^{1.3}, \qquad (3)$$

where $r$ denotes the radius in the $x - y$ plane from the source. This equation yields $H_{Disk}/R_{Disk} \approx 0.1$ and $\approx 0.6$ at $r/R_{Disk} = 0.1$ and 1.0, respectively, closely matching the flared disk shape adopted in other modeling work (see, e.g., Cazzoletti et al. 2018). In addition, we set the dust fraction in the inner H I disk to the dust fraction of the ISM (van Boekel et al. 2017). The properties of the dust affect the spatial diffusion of Lyα; higher dust optical depth induces less spatially extended surface brightness, as we will discuss further in Section 4.4.3.

We also consider a rotating motion in the disk, as the disk rotation in TW Hya roughly follows a Keplerian motion, even though the disk's kinematics have small spatial variation (Teague et al. 2022). Thus, we assume the rotating velocity in the disk $v_{rot}$ is given by

$$v_{rot}(r, z) = \sqrt{\frac{GMr^2}{(r^2 + z^2)^{3/2}}} = v_{Kep}\sqrt{\frac{r^2}{(r^2 + z^2)^{3/2}}} \qquad (4)$$

, where $r = \sqrt{x^2 + y^2}$ and $z$ are the radius and height of the disk, respectively. Assuming $r$ and $z$ in a unit of au,

$$v_{rot}(r, z) \sim v_{Kep}\sqrt{\frac{r^2}{(r^2 + z^2)^{3/2}}}, \qquad (5)$$

where $v_{Kep}$ is a Keplerian velocity, $\sim 30$ km s$^{-1}$ at $M = M\odot$. However, the rotating motion does not affect the formation of the Lyα spectrum, as the line of sight is perpendicular to the rotating velocity (Garavito-Camargo et al. 2014; Remolina-Gutiérrez & Forero-Romero 2019). Furthermore, in our simulation, the scale of the rotating motion of the protoplanetary disk (30 km s$^{-1}$) is much smaller than the width of observed Lyα emission in Figure 2 ($> 500$ km s$^{-1}$). Thus, the rotating motion in our model does not affect the Lyα radiative transfer. In Section 5.2, we will discuss the effect of changing dust properties and rotating velocity in multiple lines of sight, not only face-on direction.

### 4.3  Formation of Lyα Spectrum through a Bipolar Wind

The Lyα spectrum is mainly determined by the physical properties of H I medium in the line of sight (Herczeg et al. 2004; Arulanantham et al. 2023). Because TW Hya is a face-on object, understanding the properties of the bipolar wind is crucial in analyzing its Lyα spectrum. Hence, we employ the wind-only model to fit the central Lyα spectrum, which corresponds to the spectrum with 0″ spatial offset in the bottom panel of Figure 2. To facilitate direct comparison, we extract the simulated Lyα spectra using a slit-like detector with a width of 10 au ($\sim 0.2″$) and a 10 au bin of spatial offset (see Appendix B for details).

Figure 5 showcases the simulated results of the best model and their concurrence with the observed data. In the left panel, the spectrum of the best model is similar to the observed spectrum, except for a minor deviation in the blueward region. The best model parameters are as follows: H I column density $N_{HI} = 10^{20}$ cm$^{-2}$, expansion velocity $v_{exp} = 200$ km s$^{-1}$, and intrinsic Lyα emission width from the central source $\sigma_{src} = 300$ km s$^{-1}$. Determination of the best model involved exploring parameter ranges, as discussed in Section 4.2.1.

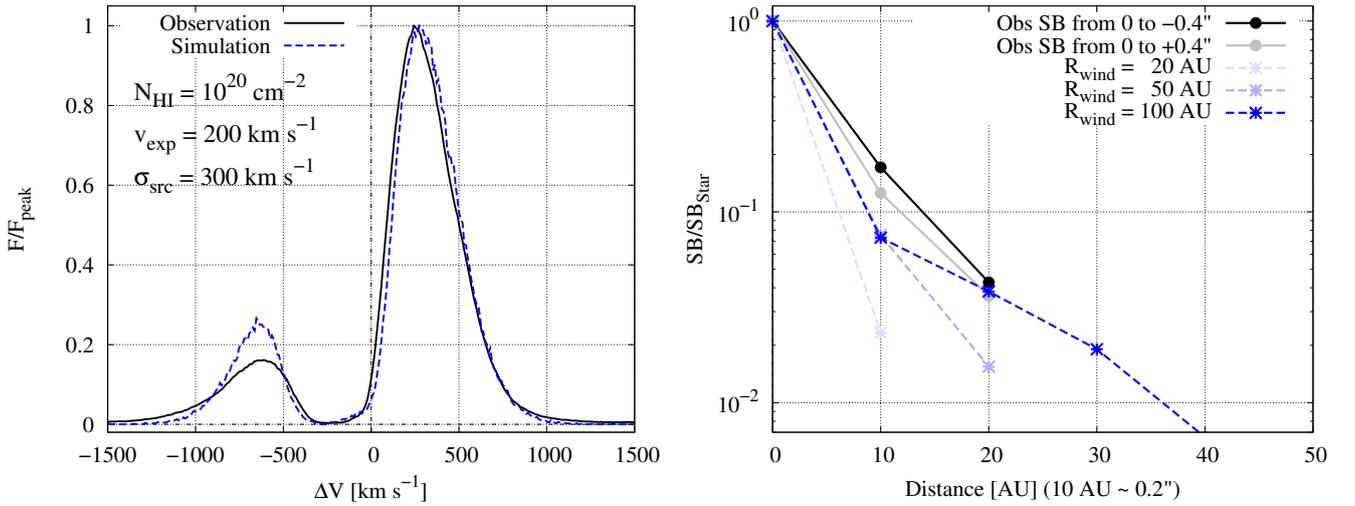

**Figure 5.** Comparisons between the observation and the simulated results obtained from the best wind-only model, focusing on the Lyα spectra with 0″ spatial offset (left) and the slit surface brightness (right). The best model parameters are determined as follows: a H I column density of the wind $N_{HI} = 10^{20}\,cm^{-2}$, an expansion velocity $v_{exp} = 200\,km\,s^{-1}$, and a width of the intrinsic Lyα emission from the central source $\sigma_{src} = 300\,km\,s^{-1}$. *Left:* The blue dashed line corresponds to the simulated Lyα spectrum of the best model. The black solid line represents the Lyα spectrum integrated with five spectra in the bottom panel of Figure 2. *Right:* The blue dashed lines indicate the simulated slit surface brightness from the simulation for various wind radii $R_{Wind}$. The black and gray solid lines illustrate the profiles of the slit surface brightness at positive and negative offsets, respectively. Each point represents the total integrated flux of the 2D spectrum in Figure 1.

Appendix C provides additional insights into the behavior of the spectral shape with varying parameters, revealing that the Lyα spectrum profile is predominantly governed by $N_{HI}$, $v_{exp}$, and $\sigma_{src}$ (see Figure C1). Notably, the wind radius $R_{Wind}$ and wind temperature $T_{Wind}$ have negligible impact on the Lyα spectrum profile (see Figure C2).

However, while the wind-only model successfully explains the 1D Lyα spectrum, it encounters challenges in explaining the spatial distribution of Lyα. The right panel of Figure 5 illustrates the slit surface brightness as a function of the projected radius $R_p$ for various wind radii $R_{Wind}$. To calculate the slit surface brightness profile from the simulation, we compute the total flux in each slit (see details in Appendix B). Although the simulated profile extends further with increasing $R_{Wind}$, it exhibits a steeper profile compared to the observed profile. Consequently, our findings indicate the necessity of an additional scattering medium to explain the observed surface brightness profile. Therefore, in the following section, we consider the disk component in the wind-disk model.

### 4.4 Spatially Resolved Lyα in Wind-Disk Model

In the dense H I region of a protoplanetary disk, scattering allows Lyα to exhibit spatial extension. Since the disk surrounding TW Hya is observed face-on, perpendicular scattering is required to reproduce the Lyα observations. While dust scattering can contribute to spatial diffusion, its signature is hard to detect in face-on systems as it has a preference for forward scattering along the radial direction of the disk (Henyey & Greenstein 1941) Instead, Lyα scattering with atomic hydrogen is required. Accordingly, this study focuses on the H I component in the inner disk, characterized by $R_{Disk,HI}$, as discussed in Section 4.2.2, to model the Lyα surface brightness observed with *HST*-STIS. Moreover, we explore the behavior of the offset Lyα spectra by varying the radius of the inner H I disk $R_{Disk,HI}$.

#### 4.4.1 Spatial Diffusion through Disk Scattering

Figure 6 shows the slit surface brightness profiles for various $R_{Disk,HI}$. A clear trend emerges, indicating that the profile becomes more extended with increasing $R_{Disk,HI}$. Specifically, at $R_{Wind} = 50$ and 100 au, the profiles with $R_{Disk,HI} \gtrsim 20$ au closely match or are even more extended than the observed surface brightness. This finding leads us to conclude that scattering with atomic hydrogen in the inner H I disk induces spatial diffusion, effectively explaining the observed surface brightness when $R_{Disk,HI} \gtrsim 20$ au.

#### 4.4.2 Offset spectrum

In Figure 7, we compare the simulated and observed Lyα spectra for offsets ranging from $0 - 4''$ for various $R_{Disk,HI}$ values. All observed spectra exhibit distinct doublet peaks, featuring an enhanced red peak and a blue absorption feature in the Doppler factor at $\sim -100\,km\,s^{-1}$. The simulated spectra in the redward of the line center show weak dependence on $R_{Disk,HI}$ and closely resemble the observations. However, the spectra in the range of $\Delta V = -200 - 0\,km\,s^{-1}$ become increasingly enhanced with increasing $R_{Disk,HI}$, especially $R_{Disk,HI} > 0.5R_{Wind}$ (see dotted lines). This enhancement arises from scattering in the disk region not covered by the wind. A schematic illustration of this effect is shown in Figure 8.

When the wind obscures the H I inner disk, the intrinsic photons in the blueward ($\Delta V = -v_{exp} - 0$) must undergo scattering in the wind. As a result, the escaping spectrum exhibits absorption features in the blueward. However, when the size of the wind is smaller, intrinsic photons in this $\Delta V$ range can escape to the line of sight without interacting with the wind via scattering in the inner H I disk. Consequently, the escaping spectrum appears similar to the intrinsic Gaussian profile, as shown in the dotted lines in the top right panel of Figure 7. To effectively explain the spatially resolved Lyα spectra, we conclude that the wind must be sufficiently large to cover the inner H I disk, that is, $\sin\theta_{o,Wind} R_{Wind} > R_{Disk,HI}$.

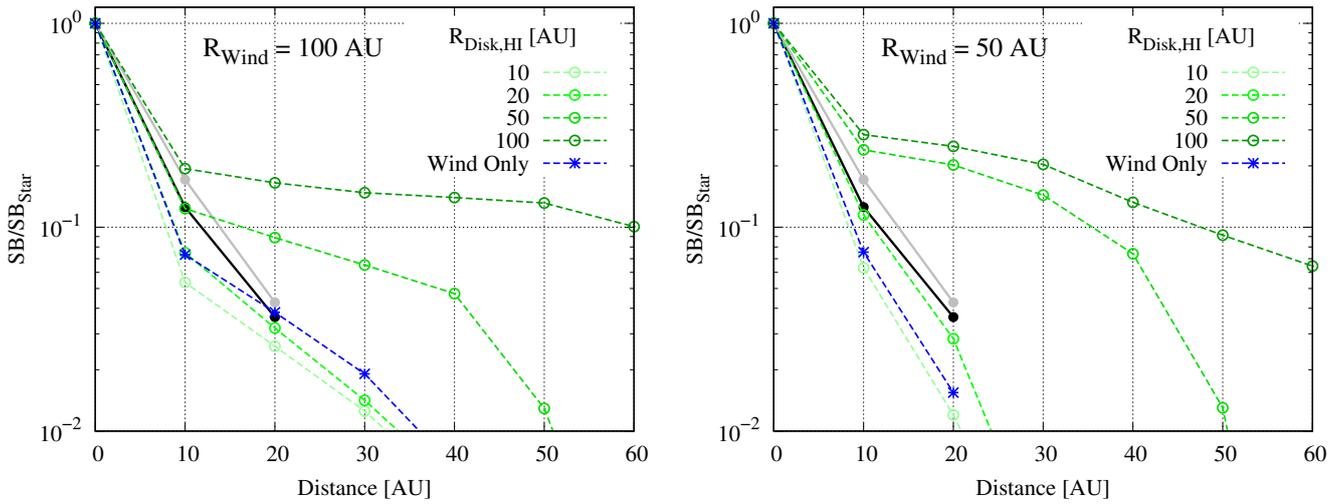

**Figure 6.** Slit surface brightness profiles for various $R_{Disk,HI}$ values at $R_{Wind} = 100$ (left) and 50 au (right). The shades of green dashed lines represent different $R_{Disk,HI}$ values ranging from 10 to 100 au. The blue dashed lines show the surface brightness of the wind-only model corresponding to the profiles in the right panel of Figure 5. The black and gray lines represent the observed surface brightness profiles in the positive and negative offsets, respectively, as shown in Figure 5.

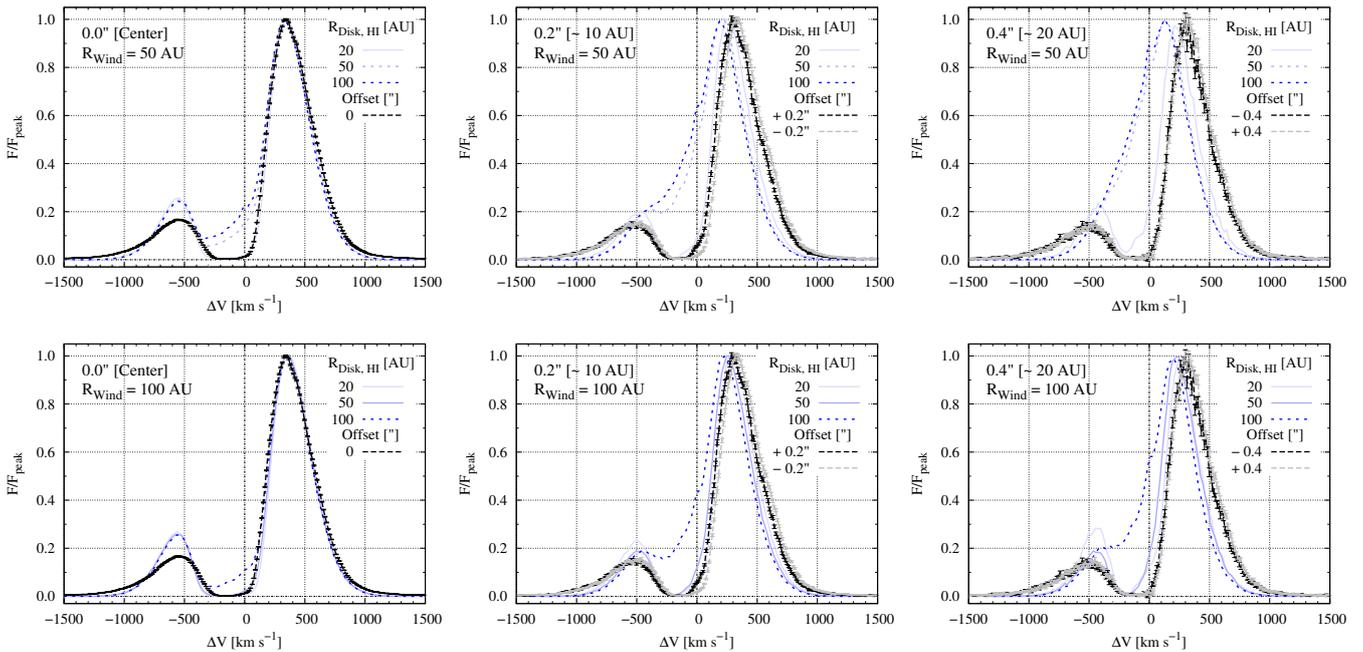

**Figure 7.** Slit spectra with spatial offsets of $0''$ (center; left), $0.2''$ ($\sim 10$ au; center), and $0.4''$ ($\sim 20$ au; right) for various $R_{Disk,HI}$ ranging from $20 - 100$ au. The top and bottom panels show the spectra at $R_{Wind} = 50$ and 100 au, respectively. The black and gray dashed lines represent observed spectra as shown in the bottom panel of Figure 2. The shades of blue colors represent different $R_{Disk,HI}$. The solid (dotted) blue lines correspond to the spectra when $R_{Disk,HI}$ is smaller (larger) than the wind-radius projected in the $x$-$y$ plane, $0.5R_{Wind} = \sin\theta_{o,Wind} R_{Wind}$. When the bipolar wind fully covers the inner H I disk, the offset spectra from the simulation provide us with a profile well-fitted with the observations.

### 4.4.3 2D spectrum

Figure 9 illustrates the simulated 2D spectra for three spatial offsets: 0, 10, and 20 au. With increasing offsets, the 2D spectrum displays greater spatial extension, akin to the observed 2D spectrum shown in Figure 1. For a more detailed analysis, Figures 10-11 facilitate a direct comparison between simulated and observed data through the 1D spectrum and the radial surface brightness extracted from the 2D spectrum.

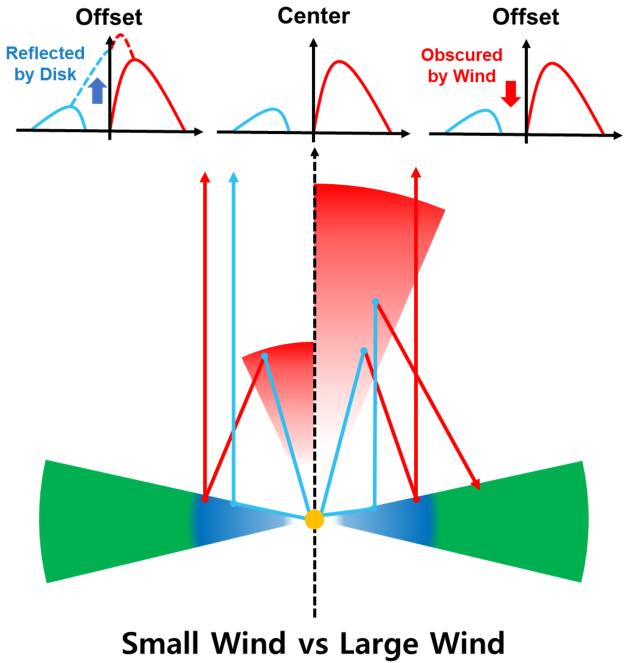

**Figure 8.** Schematic illustration explaining the behavior of Lyα spectra depending on the wind radius. The red, blue, and green regions represent the outflowing wind, inner H I disk, and outer dusty disk, as shown in Figure 4. The blue and red arrows depict the paths of photons in the blueward and redward, respectively. Because of the outflowing wind, photons in the blueward, specifically $\Delta V = -v_{exp} - 0$, experience scattering in the wind and then become redshifted. The photons in the redward can pass through the wind without significant scattering. Thus, the 'Center' Lyα spectrum, which is obscured by the wind, always exhibits clear doublet peaks. On the right side, the 'Offset' spectrum also shows a clear doublet-peaked profile when the wind covers the entire H I disk (see the blue solid lines in Figure 7). Conversely, on the left side, when the wind is small, it allows the escape of blue photons reflected by the H I disk (see the blue dotted lines in Figure 7).

In Figure 10, we compare the information extracted from the observed and simulated 2D spectrum. The left panel shows the observed 1D spectra with 0.875″ offset extracted from the 2D spectrum on the star and the simulated 1D spectra with ∼ 45 au offset (0.875″ ∼ 52.5 au, assuming 0.2″ = 12 au at the source distance of 60 pc). The simulated spectrum for $R_{Wind} = 100$ au and $\theta_{o,Wind} = 30°$ appears narrower than the observed offset spectrum at ±0.875″, and the red peak is closer to the systemic velocity. In our scattering geometry, as shown in Figure 4, the bipolar wind takes on the shape of a cone with an opening angle $\theta_{o,Wind}$. In this geometry, intrinsic emission undergoes scattering in the H I medium with the range of outflow velocity from $0.2v_{exp}$ to $v_{exp}$ in the line of sight. On the other hand, when the photon scattered by the outer H I disk positioned in alignment at the wind's edge is observed along the line of sight, the wind only covers outflow velocities in the vicinity of $v_{exp} \cos \theta_{o,Wind}$. Thus, the absorption feature of the simulated offset spectrum becomes narrow, and the red spectral peak is less shifted.

However, in the left panel of Figure 10, the 1D spectra for larger opening angle ($R_{Wind} = 100$ au & $\theta_{o,Wind} = 45°$) and extended $R_{Wind} = 200$ au & $\theta_{o,Wind} = 30°$ are closely matched with the observed spectrum near the systemic velocity. Because larger $R_{Wind}$ and $\theta_{o,Wind}$ cause a broader range of outflow velocity in the face-on direction if even Lyα is scattered by outer H I disk. In the right panel, to clarify this behavior, we estimate the fraction of core flux, which is the integrated flux of the spectrum in the range of Doppler factor from $-200$ km s$^{-1}$ to 0 km s$^{-1}$, as a function of the slit radial distance. The core fraction increases with increasing the slit distance. Particularly, when the slit distance is larger than the projected radius of the wind, the core fraction is dramatically enhanced. This behavior of the core flux allows us to consider varying wind shapes or increasing the wind size. For instance, a collimated wind with a cylindrical shape could potentially encompass a broader range of outflows in the outer H I disk.

In Figure 11, we compare the observed radial surface bright profiles for ±0.2″ and ±0.4″ offsets and simulated profiles for higher dust optical depths. The observed profile (black/gray solid lines) proves steeper than the simulated one with ISM dust fraction (black dashed line), primarily due to dust absorption. However, the protoplanetary disk consists of an H I surface along with a dense inner dusty disk (Bergin et al. 2003). To test the effect of the dense, dusty disk, in the right panel, the radial profile with the dust fraction $f_D = 100 \, f_{D,ISM}$, where $f_{D,ISM}$ is the dust fraction of ISM, is similar to the observed profile. Calahan et al. (2023) reported that the H I surface of the protoplanetary disk is less dusty than ISM. However, the broad intrinsic Lyα can penetrate the dust-free surface because of a smaller cross-section far from the line center ($< 10^{-20}$cm$^2$). As a result, Lyα radiation in the inner disk is able to be absorbed by the dust in the disk. Furthermore, we will discuss Lyα radiation within the disk in Section 5.1.

## 5 DISCUSSION

In the previous section, we presented the analysis of escaping Lyα spectra in a face-on protoplanetary disk and compared them with the observed Lyα emission from the iconic T-Tauri star TW Hya. In this section, we discuss the effects of Lyα RT in T-Tauri stars in general. In Section 5.1, we extend our investigation by computing the radiation field of Lyα within both the wind and the protoplanetary disk. This analysis allows us to explore the potential photodissociation of molecules induced by Lyα photons. Furthermore, we show that the bipolar wind plays a crucial role in enhancing the presence of Lyα photons within the disk. The scattering in the wind allows more Lyα photons to penetrate the disk. In Section 5.2, we consider various observing directions to investigate the dependence on the line of sight because the variation in observing direction causes substantial effects on the observed Lyα spectra. In Section 5.3, we explore the impact of varying density and velocity profiles in the wind-only model. Our findings suggest that different wind properties constrain higher or lower outflow velocity. In Section 5.4, we discuss the broad width of intrinsic Lyα as constrained by our modeling. We note that the width, which is about 300 km s$^{-1}$, is 3-4 times wider than the line width of Hα emission. In the following sections, we explore those aspects to emphasize the importance of Lyα RT in T-Tauri stars.

### 5.1 Internal Lyα in protoplanetary disk

The UV radiation fields generated by T Tauri stars play an important role in establishing the chemical compositions of protoplanetary disks (Aikawa & Herbst 1999; Gorti & Hollenbach 2008; Cazzoletti et al. 2018). Lyα emission is generally the strongest component in the spectrum of T Tauri stars (France et al. 2014) and may itself significantly impact the disk (Bergin et al. 2003; Bethell & Bergin 2011). Specifically, the photodissociation cross sections of $H_2O$ and HCN in the vicinity of Lyα are $\sim 10^{-17} cm^2$ (Heays et al. 2017), which is larger than the Lyα scattering cross section in the wing regime ($< 10^{-18} cm^2$). In disk layers where the relative abundances of $H_2O$ and HCN are larger than the H I fraction, the molecules will be dissociated by Lyα photons. For this reason, the fraction of Lyα photons passing through the H I surface is an essential input for models of protoplanetary disk chemistry. In this section, we compute the internal Lyα radiation field in the wind and disk from our simulation. We apply the method described in Section 2.5 of Seon & Kim (2020) to obtain the internal Lyα radiation.

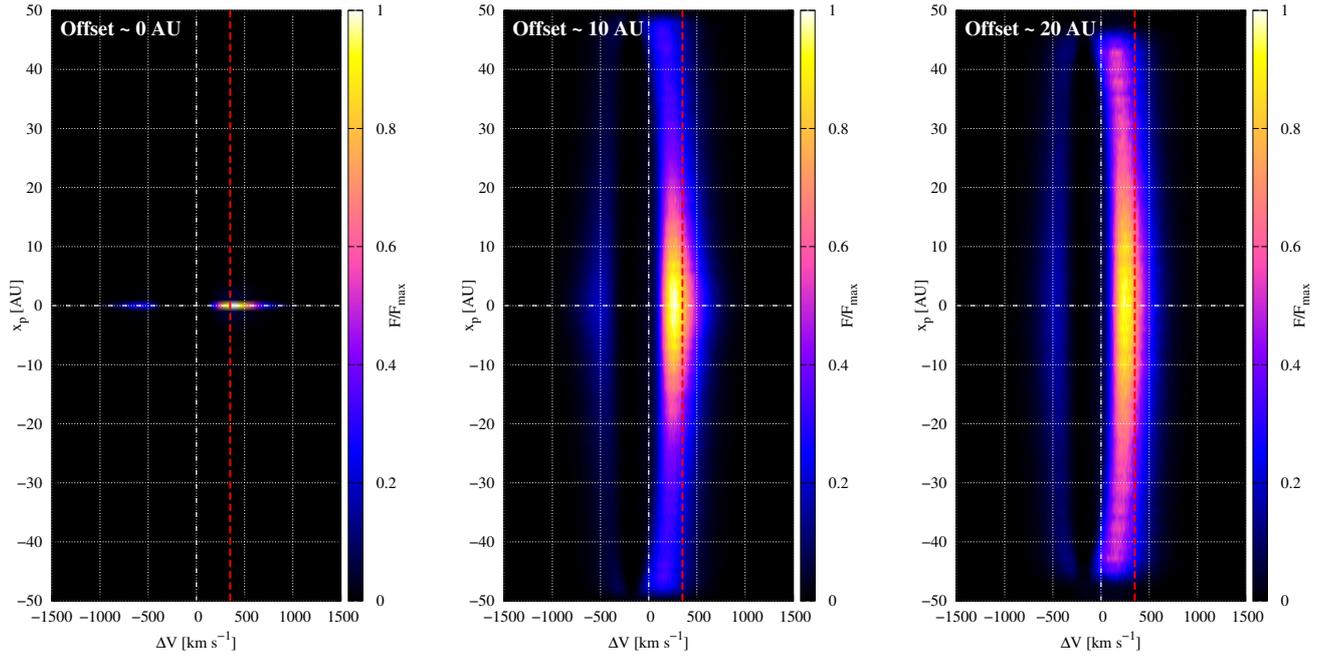

**Figure 9.** 2D spectra of the slit with spatial offsets of 0 au (left), 10 au (center), and 20 au (right), corresponding to the slits 1-3 in Figure B1. The spectra are normalized by dividing each flux by the maximum flux, similar to the HST 2D spectra shown in Figure 1. The $x$ and $y$ axes represent the Doppler factor (velocity) and the slit distance (au), respectively. The vertical red dashed line means the Doppler factor of the red peak ($\sim 250\ \mathrm{km\,s^{-1}}$) in the HST spectrum from Figure 1.

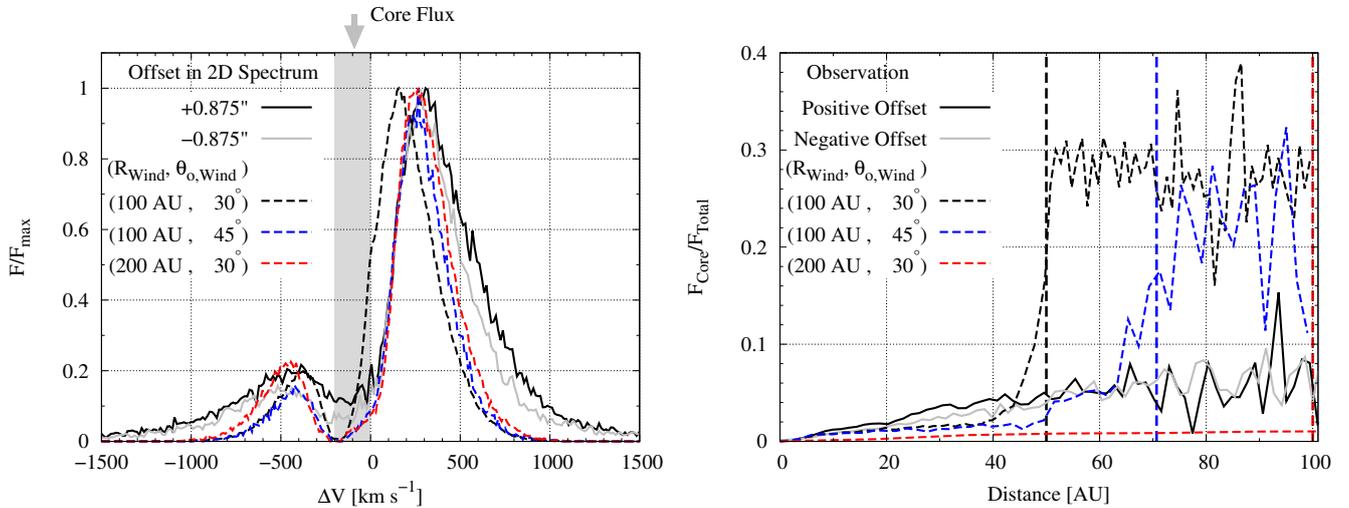

**Figure 10.** Comparison of the observation and simulation 1D spectra extracted from 2D spectra on the star to investigate the geometrical properties of the bipolar wind. *Left:* 1D extracted spectra of the observed and simulated 2D spectrum on the star in Figure 1) and Figure 9. The solid (dashed) lines represent spectra extracted from observed (simulated) 2D slit spectra, considering a width of $0.250''$ ($\sim 12.5$ au). The black/gray solid lines indicate observed 1D spectra with spatial offset $+/-0.875''$ ($\sim 45$ au) and are identical to the offset 1D spectra in Figure 2. The dashed lines represent the simulated 1D spectra with 50 au spatial offset for three parameter sets, $R_{\mathrm{Wind}} = 100$ au & $\theta_{\mathrm{o,Wind}} = 30°$ (black), $R_{\mathrm{Wind}} = 100$ au & $\theta_{\mathrm{o,Wind}} = 45°$ (blue), and $R_{\mathrm{Wind}} = 200$ au & $\theta_{\mathrm{o,Wind}} = 30°$ (red). *Right:* The fraction of core flux, the ratio of core flux and total flux, as a function of the distance of the slit. The core flux is integrated flux in the range of $\Delta V$ from -200 km s$^{-1}$ to 0 km s$^{-1}$ (gray zone in the left panel). The color vertical lines indicate the projected radius of the bipolar wind. The ratio increases with increasing the distance, especially when the distance is higher than the projected radius of the wind.

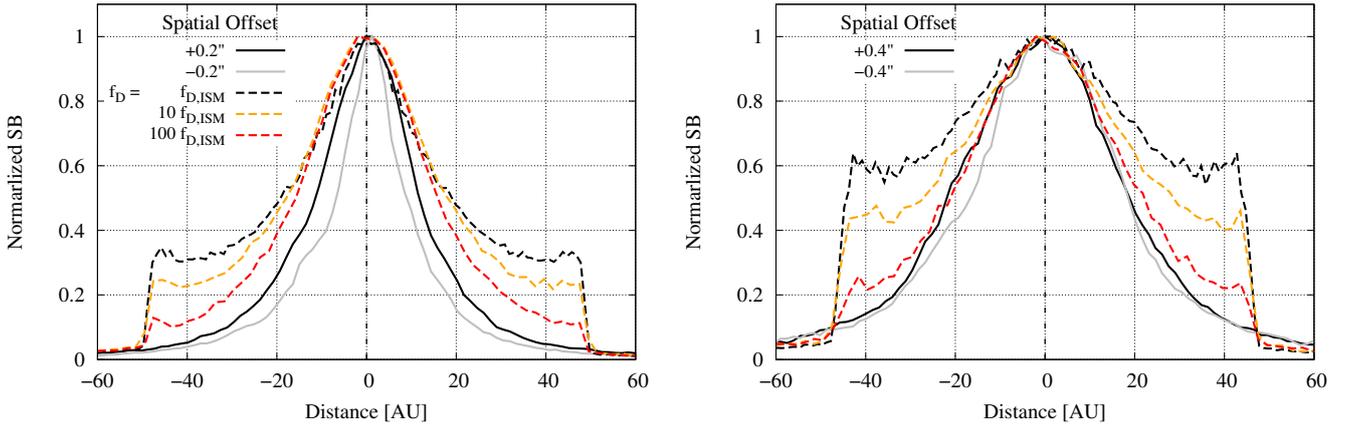

**Figure 11.** Radial surface brightness of the 2D spectra for the spatial offset ±0.2″ (left) and ±0.4″ (right). The black/gray line represents the radial profile of the observed 2-D spectrum with positive/negative offset in Figure 1. The dashed lines represent simulated profiles for various dust fractions $f_D = 1$ (black), 10 (orange), and 100 $f_{D,ISM}$ (red) where $f_{D,ISM}$ is the dust fraction of ISM. As $f_D$ increases, the simulated profile becomes similar to the observed profile due to dust extinction in the protoplanetary disk.

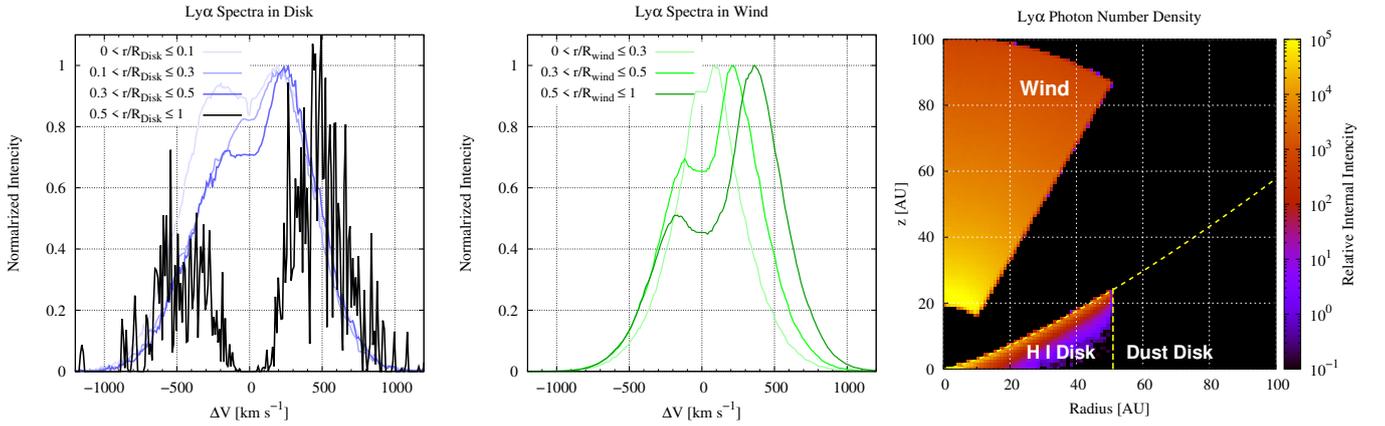

**Figure 12.** The internal Lyα radiation from our RT model with $R_{Wind} = 100$ au and $R_{Disk,HI} = 50$ au. *Left:* Normalized Lyα spectra in the disk for various radius ranges. The blue and black lines represent spectra in the inner H I disk (< 50 au) and outer dust disk (> 50 au). *Center:* Normalized Lyα spectra in the bipolar wind for various radius ranges. *Right:* the map of the internal Lyα radiation within the wind and disk per volume in the radius $r = \sqrt{x^2 + y^2}$ and $z$ plane. $R_{Wind}$ and $R_{Disk,HI}$ are fixed at 100 and 50 au, respectively.

### 5.1.1 Lyα within Wind and Disk

Figure 12 illustrates the internal Lyα radiation of the wind-disk model. In the two left panels, the spectra within the H I disk and the wind exhibit enhanced flux near the line center, while the spectrum within the dusty disk shows a significant central dip. Within the H I disk, Lyα photons near the line center undergo multiple scatterings while maintaining their wavelength due to a high optical depth. The Lyα intensity of the scattering medium is proportional to the time staying in the medium (see Eq. 25 of Seon & Kim 2020). As a result, the internal spectrum does not have a strong central dip, despite the fact that Lyα photons cannot escape at this wavelength due to the extremely high optical depth ($\tau \gg 1$; cf. Neufeld 1990). Conversely, the spectrum within the dusty disk is akin to the escaping spectrum, as the cross section of dust scattering does not strongly depend on the wavelength.

In the right panel of Figure 12, the Lyα intensity within the wind and H I inner disk ($r < R_{Disk,HI}$) substantially exceeds that within the outer dusty disk ($r > R_{Disk,HI}$). Within the dusty disk, even if Lyα photons stay for a long time through multiple dust scatterings, their intensity significantly diminishes, proportional to $\alpha^{N_D}$, where $\alpha$ and $N_{\mathrm{D}}$ denote the albedo and the number of dust scatterings, respectively. Thus, Ly$\alpha$ radiation within the outer dusty disk is much weaker than that within the inner H I disk.

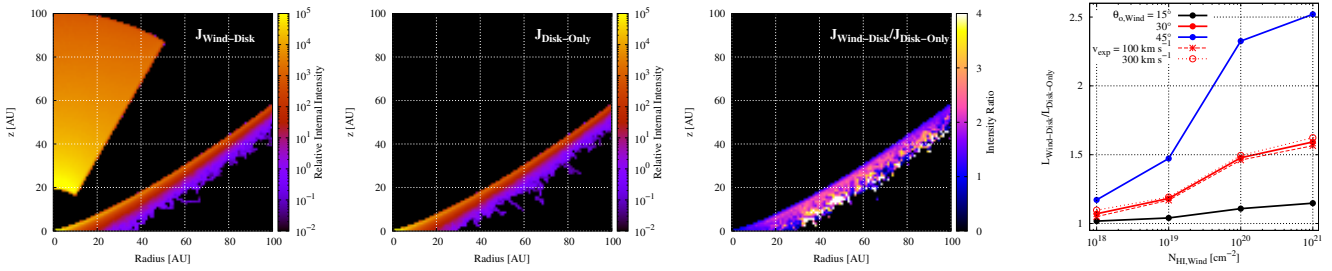

**Figure 13.** The internal Ly$\alpha$ radiation of the wind-disk and disk-only models. To study the effect of wind reflection, it is assumed that $R_{\mathrm{Wind}} = 100$ au and $R_{\mathrm{Disk, HI}} = 100$ au. The disk-only model consists of only the disk and the source. The left panels show the internal intensity maps of Ly$\alpha$ for the wind-disk and disk-only models, respectively. $N_{\mathrm{HI, Wind}}$ and $v_{\mathrm{exp}}$ are fixed at $10^{20}\,\mathrm{cm^{-2}}$ and $200\,\mathrm{km\,s^{-1}}$. The color scale of the maps is identical to that in the right panel of Figure 12. The third panel displays the ratio of the internal intensities of the wind-disk and disk-only models in the left panels. The ratio is $\sim 1$ at the disk surface but $> 2$ in the deeper disk. The right panel shows the ratio of the integrated internal radiations of the wind-disk and disk-only models as a function of $N_{\mathrm{HI, Wind}}$. The colors of solid lines represent various wind's opening angle $\theta_{\mathrm{o, Wind}}$ as $v_{\mathrm{exp}}$ is fixed at $200\,\mathrm{km\,s^{-1}}$. The red dashed and dotted lines are for $v_{\mathrm{exp}} = 100$ and $300\,\mathrm{km\,s^{-1}}$, respectively, at $\theta_{\mathrm{o, Wind}} = 30°$. The larger $\theta_{\mathrm{o, Wind}}$ and higher $N_{\mathrm{HI}}$ enhances the internal Ly$\alpha$ radiation due to the wind reflection.

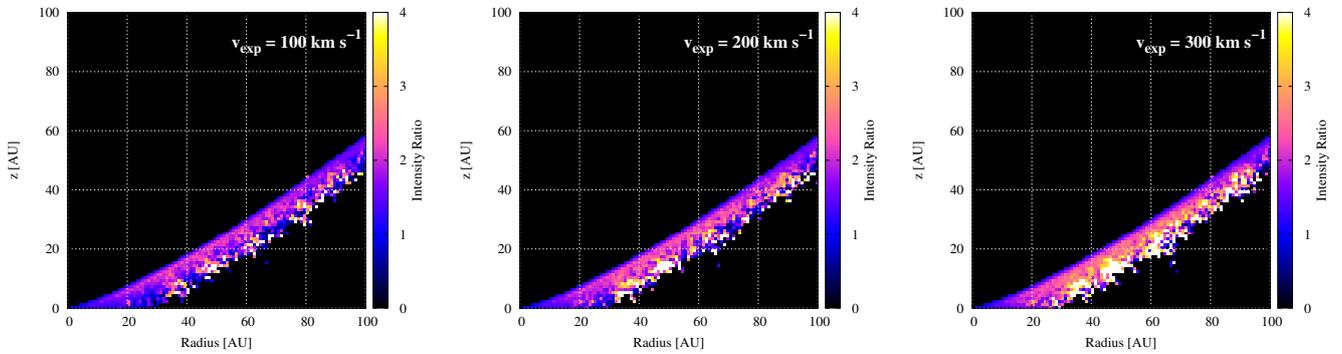

**Figure 14.** Maps of the ratio of the internal Ly$\alpha$ radiation for three $v_{\mathrm{exp}} = 100\,\mathrm{km\,s^{-1}}$ (left), $200\,\mathrm{km\,s^{-1}}$ (center), and $300\,\mathrm{km\,s^{-1}}$ (right). The map in the center panel is identical to the third panel of Figure 13; $N_{\mathrm{HI}}$ and $\theta_{\mathrm{o, Wind}}$ are fixed at $10^{20}\,\mathrm{cm^{-2}}$ and $30°$. The internal radiation in the deeper disk at $v_{\mathrm{exp}} = 300\,\mathrm{km\,s^{-1}}$ is stronger than that at $v_{\mathrm{exp}} = 100\,\mathrm{km\,s^{-1}}$.

### 5.1.2 Wind reflection of Ly$\alpha$

Through our radiative transfer modeling, we confirmed that scattering in the wind is crucial to explain observed Ly$\alpha$ spectra. Here, we discuss how wind scattering influences the internal Ly$\alpha$ radiation by reflecting Ly$\alpha$ from the star onto the protoplanetary disk. Furthermore, as the outflowing wind induces a red-shifted Ly$\alpha$, photons scattered by the wind find it easier to penetrate deeper into the disk.

In the left panel of Figure 12, the internal Ly$\alpha$ spectrum becomes symmetric with increasing the radius because of the symmetric intrinsic profile from the central source. The asymmetry of the Ly$\alpha$ spectrum in the outer region ($r/R_{\mathrm{Disk}} > 0.2$) originates from scattering in the outflowing wind. This indicates that the bipolar wind contributes to Ly$\alpha$ directed toward the polar direction, reaching the equatorial protoplanetary disk. To confirm this effect, we consider an additional geometry excluding the bipolar wind, which is the disk-only model.

In Figure 13, we compare the internal Ly$\alpha$ radiation of the wind-disk and disk-only models. The two left panels illustrate that the internal radiation within the disk of the disk-wind model is similar to that of the disk-only model. However, the reflection by the wind enhances the internal radiation, especially deep inside of the disk ($\sim 10$ au from the surface, which corresponds to an H I column density $\sim 10^{21}\,\mathrm{cm^{-2}}$). The third panel of Figure 13 shows the ratio of the internal radiation of the wind-disk and disk-only models. If the ratio is close to 1, the reflection is insufficient to affect the internal radiation. Thus, a ratio higher than 2 is observed when the reflected light is stronger than direct incident radiation from the source. Since the ratio in the deep disk is $\sim 4$, the wind reflection is crucial to understand the Ly$\alpha$ internal radiation.

To investigate the general trend varying the wind properties, the fourth panel of Figure 13 shows the ratio of the total internal radiation of the wind-disk and disk-only models. The ratio increases with increasing $N_{\mathrm{HI, Wind}}$ and $\theta_{\mathrm{o, Wind}}$. When $N_{\mathrm{HI, Wind}} \geq 10^{20}\,\mathrm{cm^{-2}}$ at $\theta_{\mathrm{o, Wind}} = 45°$, the internal radiation in the wind-disk model is two times stronger than that of the disk-only model. Consequently, the bipolar wind plays an essential role in enabling Ly$\alpha$ to contribute to the photodissociation of molecules.

In the fourth panel of Figure 13, when $v_{\mathrm{exp}}$ increases, the ratio of the total internal radiation slightly increases. However, when the photons are more redshifted through scattering in a faster outflowing medium, the scattered photons easily penetrate the H I surface due to the lower scattering cross section. Figure 14 illustrates that the internal Ly$\alpha$ radiation at $v_{\mathrm{exp}} = 300\,\mathrm{km\,s^{-1}}$ is stronger than that at $v_{exp} = 100 \ km\,s^{-1}$, especially in the deep inside of the disk. As a result, the kinematic of the wind also plays a crucial role for Lyα photon to approach the internal layers of the disk.

### 5.2 Dependence on Observing Direction

In this work, we have presented spatially resolved Lyα spectra of TW Hya, revealing consistent spectral profiles despite various spatial offsets. Additionally, our simulations have indicated that the observed Lyα profiles remain similar both on the star and with spatial offsets due to the face-on direction of the wind-disk system. While Aru-lanantham et al. (2023) reported various spectral profiles of T-Tauri stars, such as single- or double-peaked profiles with enhanced blue or red peaks, these characteristics originate from accretion and outflow of H I medium. However, since the accretion process exists near the star, achieving a significant blue shift in the offset Lyα spectrum could be challenging. Moreover, non-face-on T-Tauri stars might exhibit stronger spatial variations of Lyα emission. For example, the most intrinsic emission directly escapes from the source in the line of sight between the wind and the disk, and the source is completely obscured by the disk in the edge-on direction. Thus, in this section, we test the effect of the line of sight using our model.

To better understand the impact of the line of sight on our model, we explore the effect of various observing angles using our model geometry, as shown in Figure 4. Figure 15 offers insight into the projected images and slit spectra for three observing angles: $\theta_{obs} = 0°$ (face-on), $\theta_{obs} = 45°$ (between the wind and disk), and $\theta_{obs} = 90°$ (edge-on).

In the left panels of Figure 15, the projected image is circular and symmetric in the face-on direction ($\theta_{obs} = 0°$). The spectra of the northern and southern slits with identical offsets are similar to those of the central slit. In the center panels for the $\theta_{obs} = 45°$ case, the spectrum of the slit C exhibits symmetry and a Gaussian-like profile due to the predominance of directly escaping photons. On the other hand, the offset spectra appear asymmetric, a result of photon scattering by the wind. The southern offset spectrum has a slightly red-enhanced peak due to the disk reflection of photons scattered by the wind. In the edge-on direction ($\theta_{obs} = 90°$), only photons scattered by the winds can escape. In the right panels, the two offset spectra are similar and exhibit a red-asymmetric profile, while the center slit is completely obscured by the dusty disk.

In addition, we investigate the impact of varying rotating motion and dust properties on the Lyα profile. In Appendix D, we demonstrate that the Lyα profile is not affected by the Keplrian velocity $v_{Kep}$ and dust albedo (see Figures D1 and D2). We also varied the dust fraction of the scattering medium and found that it is still an insignificant parameter in the face-on direction. However, at $\theta_{obs} = 45°$ and high dust fraction ($f_d = 10f_{ISM}$), the spectra with spatial offset are narrower compared to those of the original model ($f_d = f_{ISM}$), since most observed photons in this line of sight experience scattering on the disk (cf. Figure D3).

Given the complex interplay between wind, disk, and line of sight, it is evident that further spatially resolved Lyα observations are necessary to investigate the structure of atomic hydrogen in T-Tauri stars. Such observations can provide valuable insights into the distribution and kinematics of H I medium and contribute to an understanding of the roles played by various factors in shaping Lyα emission profiles in T-Tauri stars.

### 5.3 Lyα profiles depending on wind properties

The kinematics and structures of a neutral medium are imprinted on Lyα emission due to its resonance nature. In other words, Lyα radiative transfer is sensitive to the physical properties of atomic hydrogen. In Section 4.2, we introduced a simplified wind model to understand the formation of Lyα profile. However, this model does not accurately represent the wind structure in all T-Tauri stars. Therefore, in this section, we investigate the effect of Lyα radiative transfer by considering different velocity and density structures in the wind. In the wind-only model, we assumed a constant density and Hubble-like outflow (Section 4.2.1). We explore two wind models - the decreasing density and the constant outflow velocity models - and compare them with our original model.

In the decreasing density case, we assumed that the density is inversely proportional to $r^2$ due to mass conservation. The H I number density $n_{HI}$ is given by $n_{HI}(r) = n_0/r^2$, where $n_0$ is a constant. The outflow velocity is still proportional to $r$, like our original model. For the constant outflow velocity model, the radial velocity of the wind $v(r)$ is always $v_{exp}$ regardless of the radius; The H I density is also constant.

In Figure 16, we show the spectra of the decreasing density and constant velocity cases and compare them to our original wind-only model. The decreasing density case shows a stronger blue peak than the best model in the wind-only model when $v_{exp} = 200 \ km\,s^{-1}$, due to the low H I density in the outer wind. On the other hand, the spectrum of the constant outflow velocity case has a more subdued blue peak at $v_{exp} = 200 \ km\,s^{-1}$ due to the constant velocity of the entire wind. In conclusion, to reproduce the simulated spectra comparable to the best model in Figure 5, the decreasing density and constant velocity cases require higher and lower $v_{exp} = 300 \ km\,s^{-1}$ and 100 $km\,s^{-1}$, respectively. In our future work, we plan to adopt more realistic wind geometries through observational constraints and theoretical expectations but note that the here employed 'effective' model can already show some of the main effects – and in fact, often radiative transfer results through more complex geometries can be mapped back to some simplified scenario (Li & Gronke 2022).

### 5.4 Broad intrinsic Lyα

Lyα and Hα lines are prominent emission lines in the UV and optical spectra, respectively, both originating from atomic hydrogen. Intuitively, the expectation is that the intrinsic line widths of these two lines are similar. However, in the case of TW Hya, the observed width of Hα (Muzerolle et al. 2000; Dupree et al. 2014; Herczeg et al. 2023), of which the full width at half maximum $\sim 200 - 300 \ km\,s^{-1}$, is 2-3 times smaller than the width of the intrinsic Lyα estimated by our RT modeling, $\sim 700 \ km\,s^{-1}$ ($\sigma_{src} = 300 \ km\,s^{-1}$). This broad Lyα is not unique to TW Hya; Arulanantham et al. (2023) reported similar narrower Hα lines compared to Lyα in other T-Tauri stars. In addition, the observed Paschen and Brackett lines are again much narrower than Hα (Wilson et al. 2022). To explain this discrepancy, they successfully reproduced the Lyα spectrum using a shell model while adopting the intrinsic width of Lyα as the Hα width. The shell model covers the whole sky from the central source, causing Lyα photons to undergo significant scattering processes and thus broadening the line profile.

However, in the geometry of our RT modeling, Lyα photons have the opportunity to escape between the wind and the disk. Consequently, our model does not lead to significant line broadening by scattering. Therefore, we need additional scattering processes or new physical phenomena to understand the broadening. We suggest two

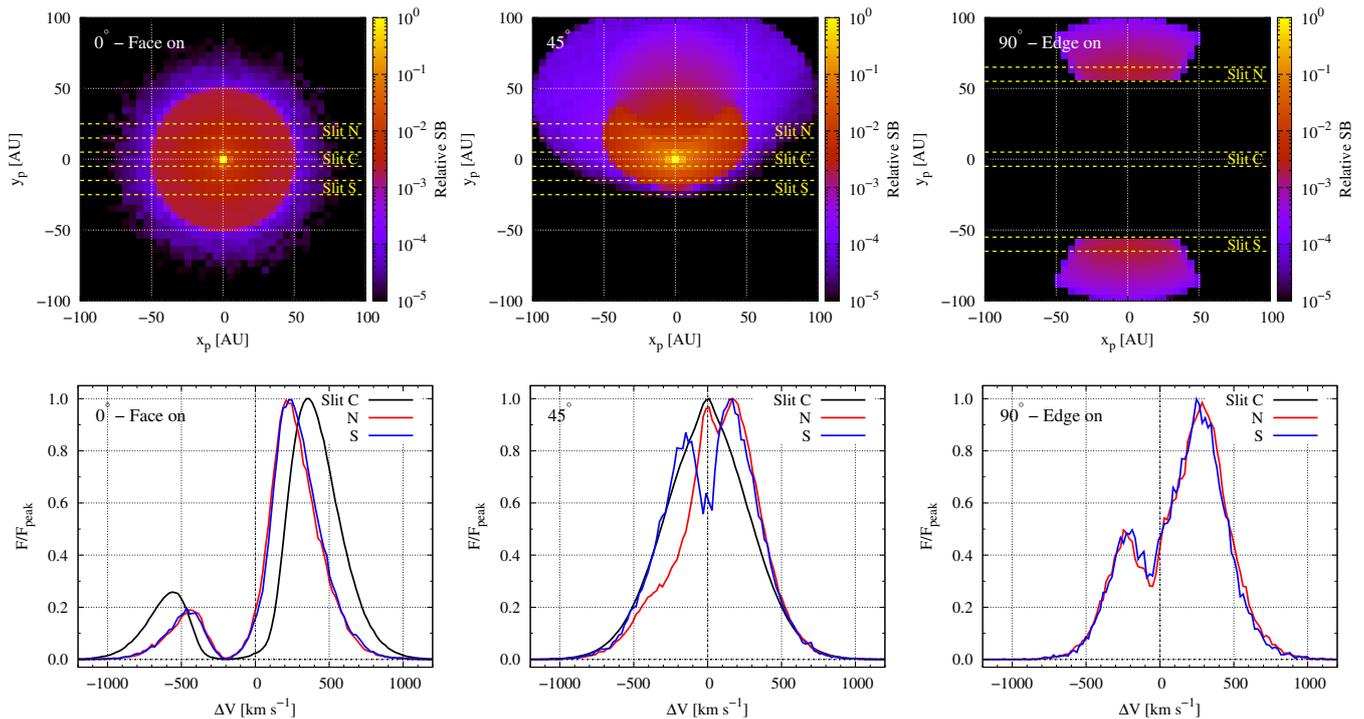

**Figure 15.** Simulated Lyα emission for three different observing angles: $\theta_{obs} = 0°$ (left), 45° (center), and 90° (right). Parameters $R_{Wind}$ and $R_{Disk,HI}$ are fixed at 100 au and 50 au, respectively. In the top panels, projected images are shown for the three observing angles, with dashed arrow lines indicating the positions of three slits (Slit C, N, and S) used to extract 1D spectra. The bottom panels depict the 1D spectra corresponding to each of the three slits: Slit C (center) in black, Slit N (north) in red, and Slit S (south) in blue. The spatial offset for Slit N and S is 20 au for $\theta_{obs} = 0°$ and 45°, and 60 au for $\theta_{obs} = 90°$.

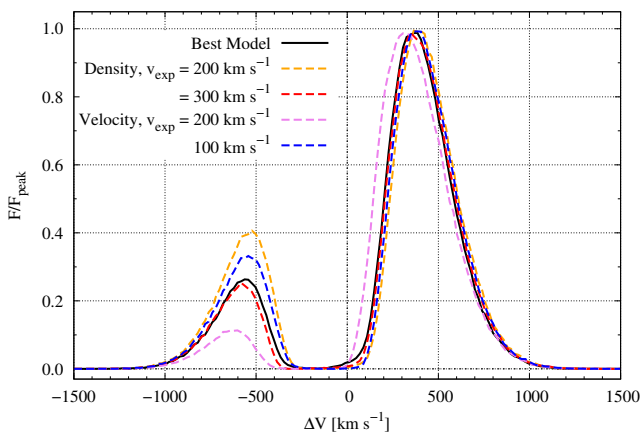

**Figure 16.** Simulated spectra of wind models with decreasing density (orange-red) and constant outflow velocity (violet-blue) cases. $N_{HI}$ and $R_{Wind}$ are fixed at $10^{20}$ cm$^{-2}$ and 100 au, respectively. The black solid line is the spectrum of the best model in Figure 5. The orange and red dashed lines represent the spectra of decreasing density case at $v_{exp} = 200$ and 300 km s$^{-1}$, respectively. The violet (blue) dashed line is the constant outflow velocity case spectrum at $v_{exp} = 200$ (100) km s$^{-1}$.

scenarios: scattering in the inner structure and Lyα from collisional excitation.

As Arulanantham et al. (2023) demonstrated, if a sufficient amount of H I medium completely surrounds the vicinity of the Lyα source,

Lyα photons must undergo numerous scattering, leading to broadening. However, the shell model tends to produce a broad double-peaked profile. Scattering in a clumpy medium prevents the formation of a double-peaked profile. A clumpy H I medium with high random motion allows us to obtain a Gaussian-like broadening of Lyα (Gronke et al. 2017; Chang et al. 2023).

The line ratio of Lyα and Hα is known to depend on the underlying physical mechanisms. The ratio is $\gtrsim 100$ due to collisional excitation (Osterbrock & Ferland 2006) and $\sim 10$ under the assumption of 'case B' recombination (Storey & Hummer 1995). Hence, if collisional excitation and recombination mainly contribute to the formation of Lyα and Hα, respectively, it is expected that the line profiles of these two lines will differ. Notably, Herczeg et al. (2023) reported the presence of a broad wing near Hα extending over 2000 km s$^{-1}$ in TW Hya. If this wing feature originates from collisional excitation, it could provide an explanation for the broad intrinsic width of Lyα observed in our modeling.

## 6 CONCLUSIONS

The dominant UV radiation in T Tauri stars is Lyα emission, which affects the chemical composition of the protoplanetary disk. The Lyα spectrum and spatial distribution depend on the physical properties of the wind and protoplanetary disk via scattering processes. This work presents spatially resolved Lyα spectra with various spatial offsets in TW Hya obtained by HST and investigates it via Lyα radiative transfer simulations.

Our main conclusions are the following:

- Our observation of spatially resolved Lyα 2D spectra, acquired at varying spatial offsets from the central star, $0''$, $\pm0.2''$, and $\pm0.4''$,(Section 3), reveals that the position of the spectral red peak of Lyα does not depend on the spatial variation as shown in Figures 1 and 2. Furthermore, the integrated spectra with different spatial offsets exhibit similarities to those on the central star (see the bottom panel of Figure 2).

- The characteristics of the wind, including its kinematics and physical properties, significantly impact the resulting Lyα spectral profile. (Section 4.3). We found the best match to the data using the wind-only model in the left panel of Figure 4 to reproduce the Lyα spectrum on the star in Figure 5 (also see Section C). Specifically, the optimal parameters identified are the wind's H I column density $N_{HI} = 10^{20}\,cm^{-2}$, expansion velocity $v_{exp} = 200\,km\,s^{-1}$, and intrinsic Lyα width $\sigma_{src} = 300\,km\,s^{-1}$. The parameters can be slightly affected by the changes in wind velocity and density profiles as discussed in Section 5.3. However, the wind-only model can not reproduce the spatial diffusion of Lyα (see the right panel of Figure 5).

- To explain the observed spatial distribution of Lyα emission, we consider a wind-disk model, where the disk is composed of an inner H I disk and an outer dusty disk (see Figure 4). The scattering in the inner H I disk leads to the spatial extension of Lyα photons. The simulated surface brightness profiles are consistent with or even more extended than the observed profiles when the radii of the H I disk $R_{Disk,HI} > 20$ and 50 au at the radii of the wind $R_{Wind} = 50$ and 100 au, respectively (see Figure 6).

- When the projected radius of the wind, $\cos\theta_{o,Wind}R_{Wind}$, exceeds the radius of the H I disk ($R_{Disk,HI}$), the simulated offset spectrum is akin to the observational spectrum (see Figure 7). On the other hand, when the wind does not obscure the H I disk, the Lyα emission near the line center easily escapes. It makes the spectrum without absorption feature in the blueward, which does not match the observed spectrum (see the schematic illustration in Figure 8). As a result, we conclude that the spatially resolved Lyα allows us to obtain the size of the wind and H I disk.

- Exploring the influence of Lyα within the protoplanetary disk, we have computed its internal radiation through our simulations (Section 5.1). This revealed significantly stronger Lyα emission within the inner H I disk compared to the outer dusty disk, as shown in the right panel of Figure 12. Thus, the presence of H I surface is essential for enabling Lyα photons to penetrate into the molecular disk and influence the chemistry of the disk.

- The reflection of Lyα by the wind is capable of transporting Lyα emission from the star to the disk. We examined a disk-only model, demonstrating that the outflowing wind causes significant radiation, illuminating the disk via scattering processes (see Figures 13 and 14). This suggests that the wind component cannot be neglected in studies of the disk composition.

- The similarity of Lyα spectra on the star and with the spatial offset is broken in the non-face-on direction. In the direction between the wind and disk, the Lyα spectrum has a less redshifted peak due to the direct escaping photons from the source. In the edge-on direction, only photons scattered by the wind are observable as the source is obscured by the dusty disk.

Lyα observables such as the emergent spectrum or surface brightness profile hold valuable information about the H I structure of T-Tauri systems – which is difficult to obtain by other means. In this study, we showed that the spatial variation of the spectrum holds additional information about the disk-wind interplay via scattering processes. To successfully reproduce the observables and to constrain the physical parameters of TW Hya, we used a simplified setup in this study. Furthermore, we showed the internal Lyα radiation within the scattering medium. In future work, we want to increase the complexity and realism of the H I geometry and kinematics.

## ACKNOWLEDGEMENTS

The authors appreciate the valuable feedback from the anonymous referee, which contributed to improving the content of the work. We thank Dr. Tom Bethell, PI of GO-11607, for his pioneering contributions in motivating and developing the original science case for these observations. Funding for the DPAC has been provided by national institutions, in particular, the institutions participating in the *Gaia* Multilateral Agreement. GJH is supported by general grant 12173003 from the National Natural Science Foundation of China. MG thanks the Max Planck Society for support through the Max Planck Research Group.

## DATA AVAILABILITY

This work has made use of data from the European Space Agency (ESA) mission *Gaia* (https://www.cosmos.esa.int/gaia), processed by the *Gaia* Data Processing and Analysis Consortium (DPAC, https://www.cosmos.esa.int/web/gaia/dpac/consortium).

## APPENDIX A: CHARACTERIZING THE INSTRUMENT RESPONSE TO OFF-CENTER STELLAR POINT SOURCES

We present spectral images of WD-1056-384, a white dwarf that was observed with identical configurations and slit positions as those used to observe TW Hya (see Figure A). In the images acquired at ±0.2″ and ±0.4″, diffraction rings from the stars (TW Hya or WD-1056-384) can still enter the 52″x0.2″ slit used for these observations. This additional stellar flux was removed by scaling each white dwarf spectral image to the peak surface brightness detected from TW Hya at the analogous position and subtracting the resulting PSF "model." The white dwarf spectral images also include geocoronal emission that fills the slit near Lyα line center, with a surface brightness of $\sim 10^{-12}$ erg s$^{-1}$ cm$^{-2}$ arcsecond$^{-2}$.

## APPENDIX B: SLIT SURFACE BRIGHTNESS AND SLIT SPECTRUM FROM RT SIMULATION

In order to compare the simulated spectra with the observations, we employ a slit shape detector with a width of 10 au, approximately 0.2″, corresponding to the slit width of the STIS slit spectrograph as shown in the bottom panel of Figure 2. The left panel of Figure B1 illustrates five slits with spatial offsets ranging from 0 to 40 au (equivalent to 0″ to 0.8″), as the range of offset in the observation is from 0″ to 0.4″. The center panel displays the spectra obtained from these various offsets, which will be used for comparison with the observed spectrum in Figure 7.

In the right panel of Figure B1, we calculate the slit surface brightness to estimate the spatial diffusion of Lyα. This slit surface brightness is normalized by dividing each value by the surface brightness at the central slit (Slit 1). The resulting surface brightness profile can be conveniently compared with the observed surface brightness, as shown in Figures 5 and 6. These slit spectra and slit surface brightness profiles are crucial for understanding how Lyα is spatially extended and how the wind and disk components influence the observed spectra.

## APPENDIX C: DEPENDENCE ON PARAMETERS IN THE WIND ONLY MODEL

To determine the optimal parameter set ($N_{\rm HI,Wind}$, $\sigma_{\rm src}$, $v_{\rm exp}$) for the best model, as shown in Figure 5, we compare the simulated spectrum with the observed spectrum with offset 0″. Figure C1 illustrates the variations in spectra for various parameters.

In the left panel of Figure C1, a H I column density of the wind $N_{\rm HI,Wind}$ mainly determines the red peak and wings of the spectrum. As $N_{\rm HI}$ increases, the spectrum becomes more redshifted. In the center panel, an increase in the width of the intrinsic Lyα profile $\sigma_{\rm src}$ broadens the width of the simulated spectrum. In the right panel, the blue peak shifts further with increasing expansion velocity $v_{\rm exp}$. The range of outflow velocity in the wind extends from 0.2 $v_{\rm exp}$ to $v_{\rm exp}$ due to its proportional distance from the source. This outflow covers the intrinsic spectrum from $-v_{\rm exp}$ to $-0.2v_{\rm exp}$, resulting in absorption-like features in this velocity range.

Figure C2 shows the Lyα spectra for various wind radii $R_{\rm Wind}$ and temperatures $T_{\rm Wind}$. Interestingly, these two parameters have negligible impact on the formation of the Lyα spectrum in the wind-only model. The parameter $T_{\rm Wind}$ represents the thermal motion of the scattering medium, with $v_{\rm th} = \sqrt{2kT_{\rm Wind}/m_H}$ being 4.3 and 13.7 km s$^{-1}$ at $T_{\rm Wind} = 10^3$ and $10^4$, K, respectively. In Section 4.3, we assumed high values for $N_{\rm HI} \sim 10^{20}$ cm$^{-2}$, which enable Lyα photons to escape with wavelengths far from the line center through numerous scattering events. Also, the fast outflow with $v_{exp} \sim 200\,\mathrm{km\,s^{-1}}$ causes a large Doppler shift compared to the thermal motion (Chang et al. 2023). Consequently, these factors render the dependence on temperature $T_{Wind}$ negligible.

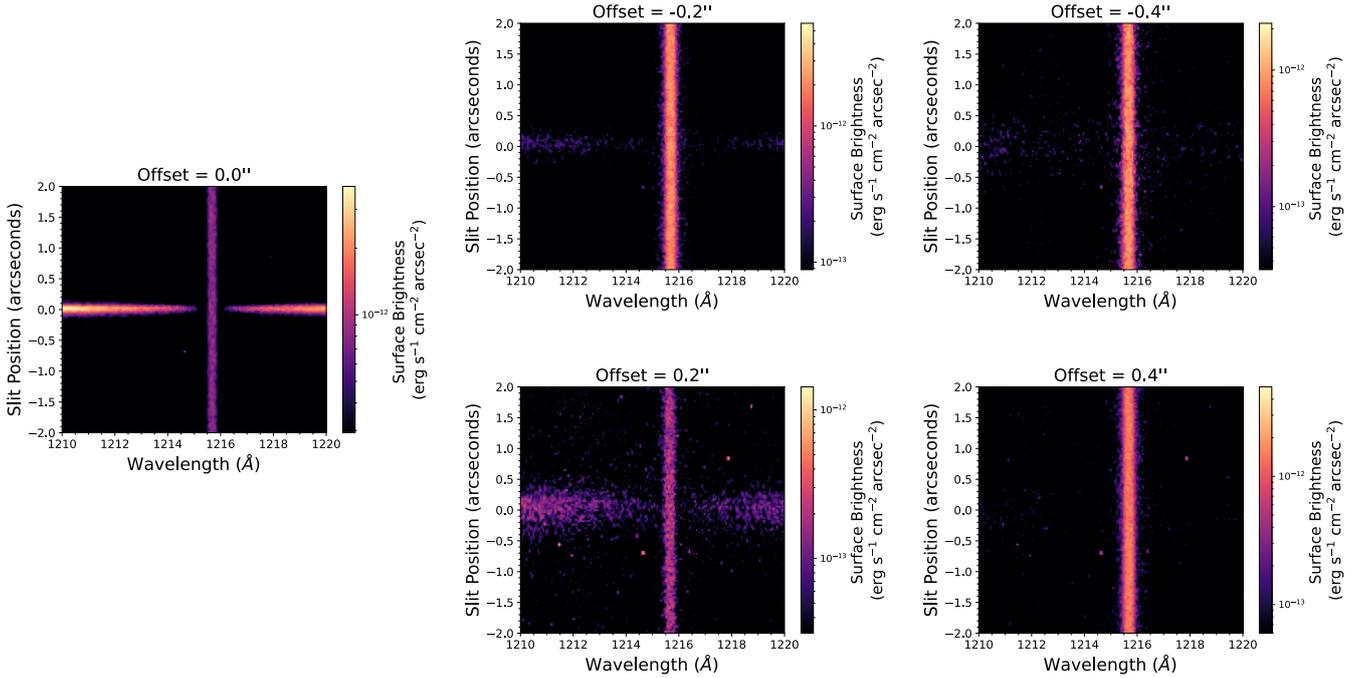

**Figure A1.** *HST*-STIS spectral images of the white dwarf WD-1056-384, acquired to characterize the PSF on the source and at offset positions +0.2″ and +0.4″. When an adjacent stellar point source is not centered in the slit, flux from the star can be spread into the spectral images and mask any Lyα emission originating at large separations. The white dwarf spectral images demonstrate the instrument response to a bright, stellar point source and quantify the flux from geocoronal emission that enters the slit at Lyα line center (1215.67 Å). We scale the white dwarf images to the peak surface brightness of TW Hya in spectral images acquired at the same positions and subtract the resulting PSF "models". The residual emission, shown in Figure 1, is spatially extended Lyα emission originating from the TW Hya system.

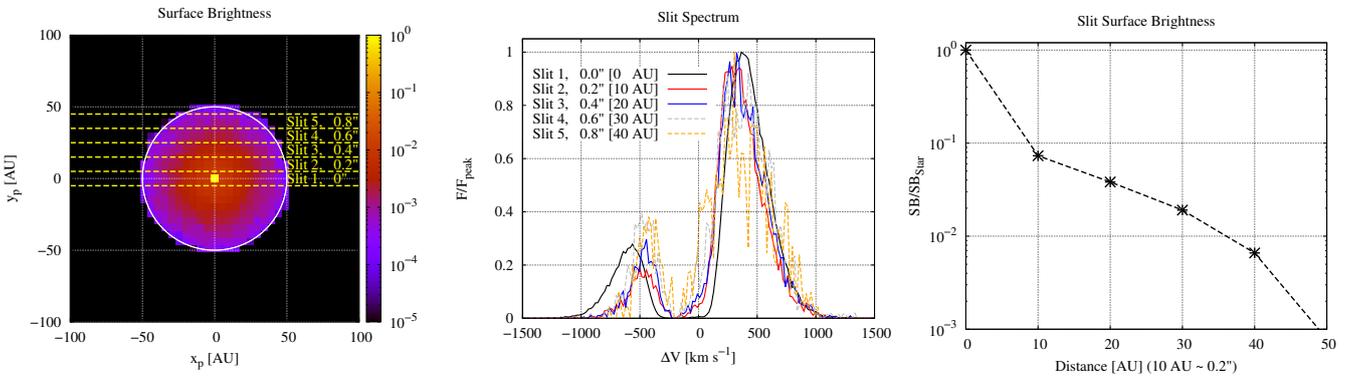

**Figure B1.** *Left:* Surface brightness distribution of Lyα in the projected $x_p - y_p$ plane. The yellow dashed lines represent the boundaries of slits with a width of 0.2″, approximately ∼ 10 au. *Center:* Spectra collected from different slits. The spectra labeled Slit 1-5 correspond to the observed spectrum with spatial offsets ranging from 0″ to 0.8″. *Right:* Slit surface brightness as a function of the distance from the central source. Each star mark represents the total number of Lyα photons in each slit divided by the number of photons in Slit 1.

## APPENDIX D: DEPENDENCE ON PARAMETERS IN THE WIND-DISK MODEL

We conducted simulations to assess the dependencies on parameters in the wind-disk model, varying Keplerian velocities $v_{Kep}$, dust albedo, and dust fraction $f_d$, in Figures D1, D2, and D3, respectively. Figures D1 and D2 clearly show that all profiles of Lyα spectra do not depend on $v_{Kep}$ and dust albedo. Furthermore, the top panel

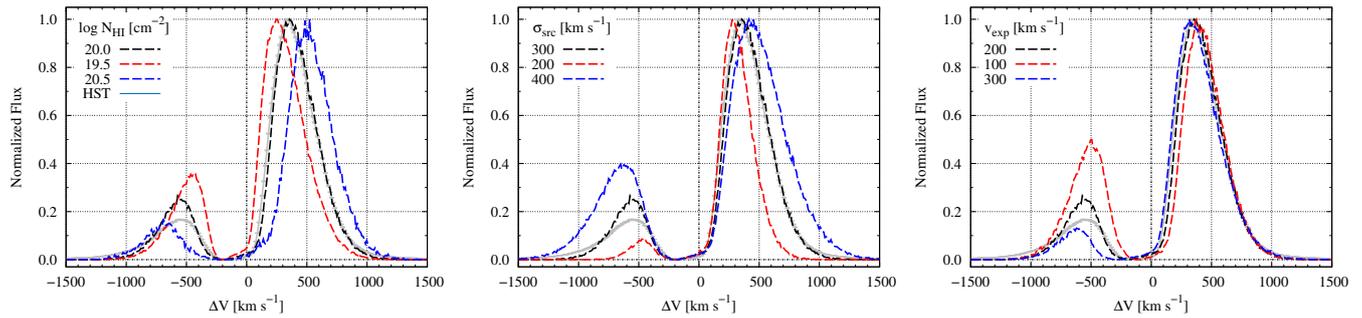

**Figure C1.** Variations of the Lyα spectrum depending on three parameters of the wind-only model: H I column density $N_{HI}$ (left), width of intrinsic Lyα $\sigma_{src}$ (center), and expansion velocity $v_{exp}$ (right). The black solid line represents the spectrum of the best model with $N_{HI} = 10^{20}\,cm^{-2}$, $\sigma_{src} = 300\,km\,s^{-1}$, and $v_{exp}$ = 200 km s$^{-1}$. The blue and red dashed lines correspond to spectra with smaller and higher values of the parameters in each panel, respectively.

This paper has been typeset from a TEX/LATEX file prepared by the author.

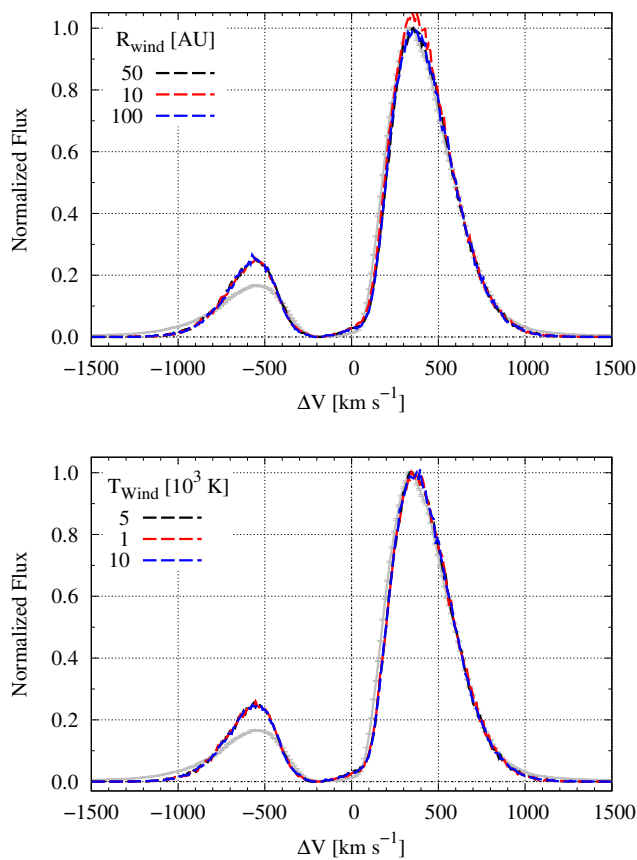

**Figure C2.** Lyα spectra varying two negligible parameters of the wind-only model: wind radius $R_{Wind}$ and temperature $T_{Wind}$. The other parameters correspond to those of the best model in Figure 5.

of Figure D3 demonstrates that $f_d$ is a negligible parameter in the face-on direction.

However, in the bottom panels of Figure D3, the spectra with $f_d/f_{d,ISM} = 10$ differ from those at $f_d/f_{d,ISM} = 1$ in the intermediate observing direction, especially the spatial offset cases (in the center and right bottom panels). This is because the red and blue wing photons primarily originate from scattering on the disk.

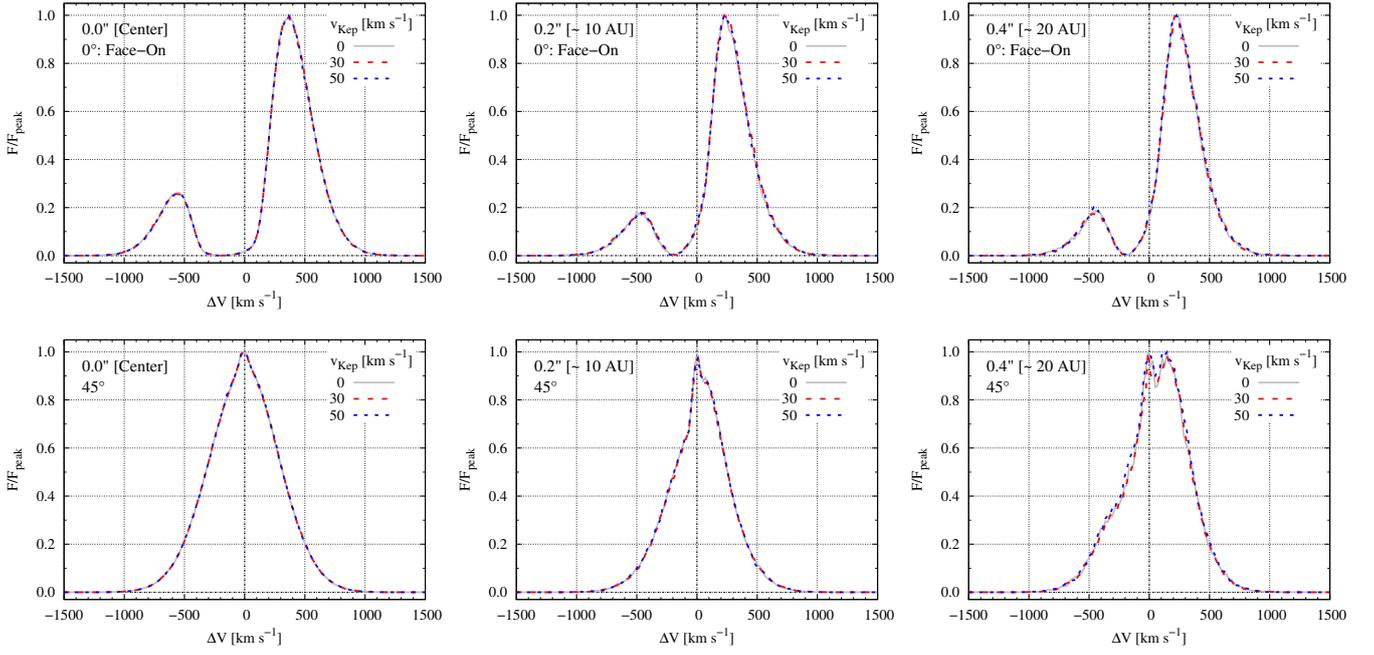

**Figure D1.** Simulated Lyα spectra for three Keplerian velocities $v_{\rm Kep} = 0\,{\rm km\,s^{-1}}$ (static), 30 km s$^{-1}$, and 50 km s$^{-1}$. The top and bottom panels show the spectra in the face-on direction 0° and intermediate direction 45°, respectively. The spatial offset is fixed at 0 au(left), 10 au(center), and 20 au(right).

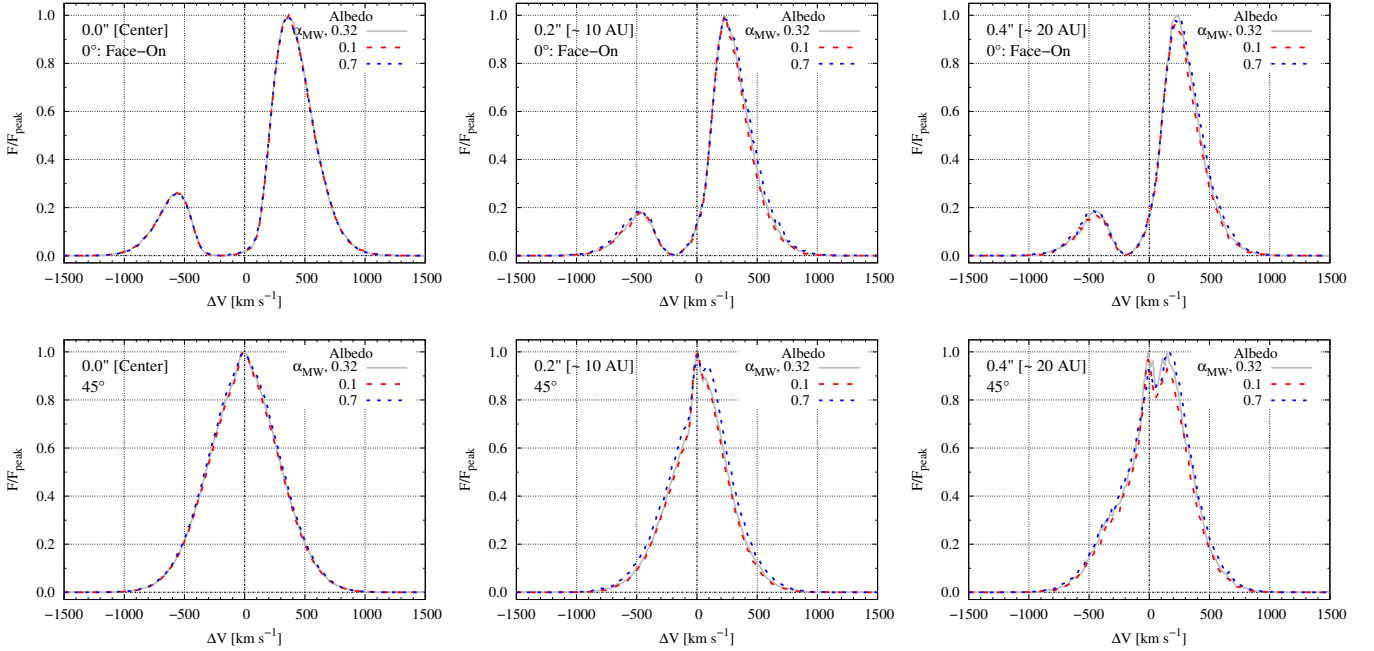

**Figure D2.** Simulated Lyα spectra for various albedo = 0.32 (MW model), 0.1, and 0.7. The structure of this figure is identical to that of Figure D1.

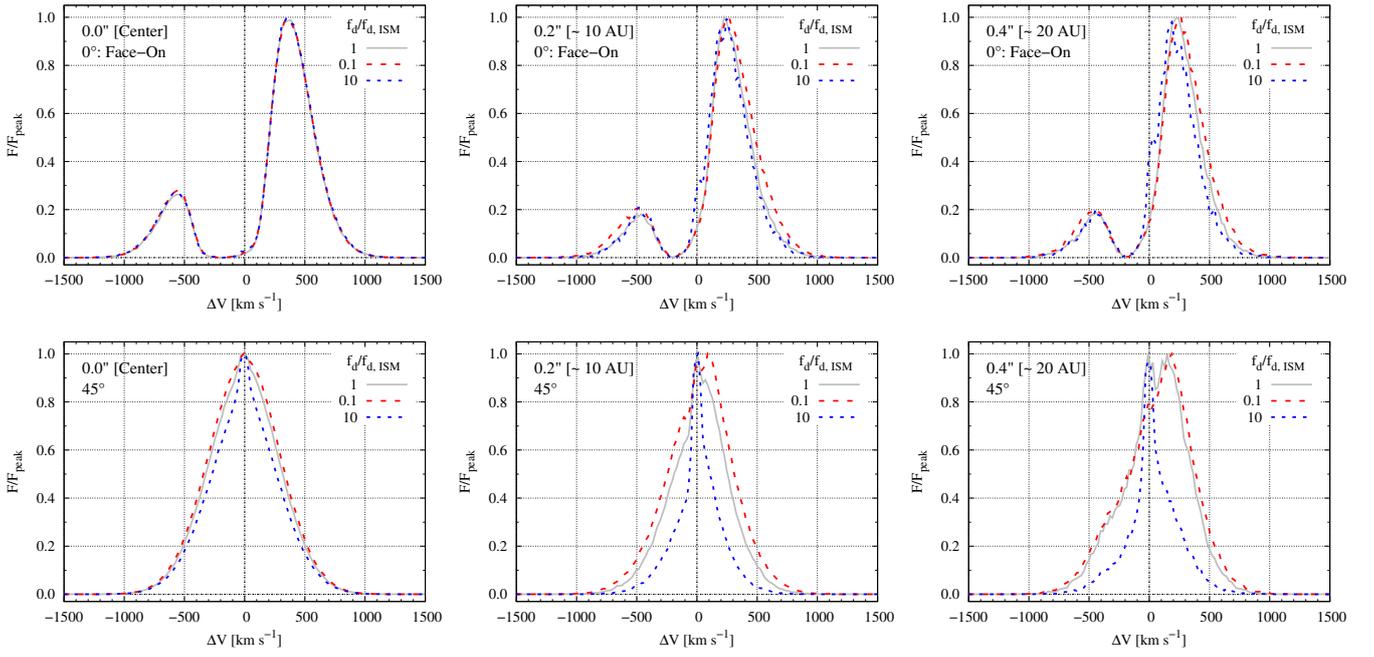

**Figure D3.** Simulated Ly$\alpha$ spectra for various dust fraction $f_d/f_{d,ISM}$ = 0.1, 1, and 10, where $f_{d,ISM}$ is the dust fraction of ISM. The structure of this figure is identical to that of Figure D1